\documentclass[useAMS,usenatbib]{mn2e}
\usepackage{graphicx}\usepackage{epsfig}\usepackage{epsf}
\usepackage{amsmath}\usepackage{amssymb}\usepackage{stfloats}
\usepackage{caption}

\title[Spectropolarimetry of SN 2009ip]{Multi-epoch spectropolarimetry of SN~2009\lowercase{ip}: direct evidence for aspherical circumstellar material}
\author[Mauerhan et al.]{Jon Mauerhan$^{1,2}$\thanks{E-mail: mauerhan@astro.berkeley.edu}, G. Grant Williams$^{2,3}$, Nathan Smith$^{2}$, Paul S. Smith$^{2}$, \newauthor Alexei V.\ Filippenko$^{1}$, Jennifer L. Hoffman$^{4}$, Peter Milne$^{2}$, Douglas~C.~Leonard$^{5}$, \newauthor Kelsey I. Clubb$^{1}$, Ori D. Fox$^{1}$, and Patrick L. Kelly$^{1}$ \\ 
 $^{1}$Department of Astronomy, University of California, Berkeley, CA 94720-3411, USA \\
 $^{2}$Steward Observatory, University of Arizona, 933 N. Cherry Ave., Tucson, AZ 85721, USA \\
 $^{3}$MMT Observatory, Tucson, AZ 85721-0065, USA \\
 $^{4}$Department of Physics \& Astronomy, University of Denver, 2112 East Wesley Avenue, Denver, CO 80208 \\
 $^{5}$Department of Astronomy, San Diego State University, PA-210, 5500 Campanile Drive, San Diego, CA 92182-1221}
\begin{document}

\pagerange{\pageref{firstpage}--\pageref{lastpage}} \pubyear{2013}
\maketitle
\label{firstpage}

\begin{abstract}
We present spectropolarimetry of SN~2009ip throughout the evolution of its 2012 explosion. During the 2012a phase, when the spectrum exhibits broad P-Cygni lines, we measure a $V$-band polarization of $P\approx0.9$\% at a position angle of $\theta\approx166^{\circ}$, indicating substantial asphericity for the 2012a outflow. Near the subsequent peak of the 2012b phase, when the spectrum shows signs of intense interaction with circumstellar material (CSM), we measure $P\approx1.7$\% and $\theta\approx72^{\circ}$, indicating a separate component of polarization during 2012b, which exhibits a higher degree of asphericity than 2012a and an orthogonal axis of symmetry on the sky. Around 30 days past peak, coincident with a substantial bump in the declining light curve, we measure $P\approx0.7$\% and another significant shift in $\theta$. At this point, broad photospheric lines have again become prominent and exhibit significant variations in $P$ relative to the continuum, particularly He~{\sc i}/Na~I~D. By 60 days past peak the continuum polarization has dropped below 0.2\%, probably declining toward a low value of interstellar polarization. The results are consistent with a scenario in which a prolate (possibly bipolar) explosion launched during the 2012a phase impacts an oblate (toroidal) distribution of CSM in 2012b.  Previous calculations that assumed spherical symmetry for the CSM have substantially underestimated the required explosion energy, since only a small fraction of the SN ejecta appears to have participated in strong CSM interaction. An ejecta kinetic energy of at least $\sim10^{51}$~ergs is difficult to avoid, supporting the interpretation that the 2012 outburst of SN~2009ip was the result of a core-collapse explosion. 
 \end{abstract}


\section{Introduction}

Interacting supernovae (SNe) are stellar explosions that collide with dense circumstellar material (CSM) 
produced by the progenitor star. These events are raising critical new questions about the final evolutionary 
phases of massive stars and the mass-loss episodes that ensue before core collapse (Smith \& Arnett 2014). 
Type-IIn and Ibn SNe, in particular, are interacting SNe that are  characterized spectroscopically by the presence 
of relatively narrow emission lines of H and He in their spectra (Schlegel 1990; Filippenko 1997; Pastorello et al. 2008), which arise from dense CSM that becomes illuminated by the shock between the fast moving SN ejecta 
and slower moving CSM (Chevalier \& Fransson 1994). As such, observations of interacting SNe probe the stellar 
progenitor's pre-SN mass-loss history, providing valuable information on its final evolutionary episodes. 

Various lines of evidence show that interacting SNe require eruptive pre-SN mass loss that is reminiscent of
luminous blue variable (LBV) stars, like $\eta$ Car, although observations indicate a wide range of 
mass-loss properties (e.g., see the review by Smith 2014). The eruptions are often detectable 
as extragalactic transients, commonly referred to as `SN impostors' (Van Dyk 2000; Smith et al. 2011a; Kochanek et al. 2012), 
which can rival the luminosity of a SN and spectroscopically mimic SNe~IIn.  Although the observational distinction between 
interacting SNe and SN impostors is not always clear, a direct link connecting these phenomena has been
 established for several objects. The progenitors of SN~2006jc and SN~2011ht were detected undergoing
 luminous outbursts 1--2~yr prior to their core-collapse explosions (Foley et al. 2007; Pastorello et al. 2008; Fraser et al. 2013a). SN~2010mc (Ofek et al. 2013a) was also detected 1--2 months before dramatically brightening, although for this case it   appears that the precursor event may actually have been the initial phases of the SN caught unusually early 
 (Smith, Mauerhan, \& Prieto 2014).
 
Perhaps the most interesting transient observed to have undergone multiple phases of eruptive mass loss is SN~2009ip. Originally classified as a SN (Maza et al. 2009), this object was actually discovered during an LBV outburst --- that is, as a SN~impostor (Smith et al. 2010; Foley et al. 2011). Several years of continued activity were followed photometrically and spectroscopically, leading up to SN~2009ip's most extreme outburst in 2012 (Mauerhan et al. 2013; Prieto et al. 2013; Levesque et al. 2014; Smith et al. 2013b; Smith, Mauerhan, \& Prieto 2014; Pastorello et al. 2013; Fraser et  al. 2013b; Margutti et al. 2014; Graham et al. 2014). The 2012 event was comprised of two main components: the ``2012a" phase, marked by an initial peak at $M\approx-15.5$ mag that lasted for just over 1 month; and the ``2012b" phase, which began with a fast 10-day rise to a second peak of $M=-18.5$ mag, followed by a bumpy decline down to a slowly declining floor in the light curve near $M\approx-13$ mag.

Based on spectral similarities with known core-collapse SNe, Mauerhan et al. (2013a) suggested that the relatively faint 2012a phase of SN~2009ip marked the initial stages of a core-collapse SN, while the subsequent 2012b brightening was the result of interaction between this SN and dense CSM ejected during the earlier LBV outbursts. Levesque et al. (2014) also shared the view that the 2012b brightening was the result of interaction between the 2012a outflow and existing CSM. However, several authors have suggested potential alternatives to a core collapse explosion -- that SN~2009ip's 2012 evolution was possibly the result of one or more nonterminal outbursts (Pastorello et al. 2013; Fraser et al. 2013b; Margutti et al. 2014), perhaps caused by the pulsational pair-instability mechanism. The motivation for nonterminal scenarios has been based largely on the fact that the total radiated energy of the light curve ($1.3\times10^{49}$ erg) can be explained by an explosion energy of $<10^{51}$~ergs (if spherical symmetry is assumed for the CSM), and also because the late-time data were interpreted as looking different  from what is expected for radioactive decay phases. More recently, Smith, Mauerhan \& Prieto (2014; hereafter SMP14) showed that the 2012 light curve and spectral evolution are consistent with published models for core-collapse SNe, which can be initially faint (like SN~1987A) owing to a relatively compact progenitor radius (i.e., a blue supergiant, as expected for an LBV, instead of a red supergiant), while the late-time characteristics could be explained if the mass of synthesized $^{56}$Ni was half that of SN~1987A. SMP14 further argued that the radiated energy did not provide an argument against core collapse, since the CSM is probably aspherical, leading to an inefficient conversion of kinetic energy into radiation. In line with the spectral modeling results and interpretation of Levesque et al. (2014), which are consistent with a toroidal distribution of CSM, SMP14 also proposed a disk-like distribution of CSM, but further argued that significant asphericity is required by the fact that broad photospheric lines are still seen at late times, even after strong CSM interaction ends.  This requires that a large fraction of the total solid angle of ejecta was able to expand unimpeded by CSM. SMP14 also pointed out that the $\sim$100-day persistence of broad photospheric lines requires a large ejecta mass of at least a few M$_{\odot}$, and is incompatible with a  0.1~M$_{\odot}$ shell, which would become optically thin much more quickly. A mass of a few M$_{\odot}$ moving at high speeds ($\sim$8000~km~s$^{-1}$) directly implies $\gtrsim10^{51}$~ergs of kinetic energy. 

In the case of SNe~IIn, the system geometry is critical if the total radiated energy from CSM interaction is to be used to infer the kinetic energy of the SN explosion. Fortunately, spectropolarimetry can yield important clues about the geometry of SNe, allowing us to stringently test the hypothesis of aspherical CSM. Here, we report multi-epoch spectropolarimetry of the 2012 evolution of SN~2009ip, from the end of the initial 2012a phase, through the peak of 2012b, and later into its decline. The results unambiguously demonstrate that the source of bright continuum emission during the 2012b phase (i.e., the CSM interaction zone) was aspherical, consistent with a toroidal or disk-like distribution of CSM proposed by SMP14 and Levesque et al. (2014). Furthermore, the results indicate that the initial 2012a phase has a separate polarization component having an axis of symmetry that is distinct from 2012b, which implies that the 2012a event did not create the CSM responsible for the 2012b brightening. Diminishing polarization at late times, in addition to similarities in wavelength-dependent polarization across lines for SN~2009ip and other SNe, provides additional evidence for a persistent underlying SN photosphere. Altogether, the available evidence supports the hypothesis of a SN explosion for SN~2009ip in 2012, and argues strongly against nonterminal explosion models for this object (at least those proposed thus far).

\section{OBSERVATIONS}
\subsection{SPOL at Arizona observatories}
Spectropolarimetric observations were obtained at the following facilities utilised by the University of Arizona: the 6.5~m Multiple Mirror Telescope (MMT) on Mt. Hopkins during 2012 Sep. 21 and 24 (UT dates are used throughout this paper); the Kuiper 61-inch telescope on Mt. Lemmon during Nov. 11 and 14; and the Bok 2.3~m telescope on Kitt Peak during Dec. 5, 6, and 7. Observations at these facilities all made use of the same CCD Imaging/Spectropolarimeter (SPOL;  Schmidt et al. 1992a). The instrument contains a rotatable semiachromatic half-waveplate used to modulate incident polarization and a Wollaston prism in the collimated beam to separate the two orthogonally polarized spectra onto a thinned, anti-reflection-coated $800 \times 1200$ pixel SITe CCD. SPOL was configured with a 600~line~mm$^{-1}$ grating in first order.  The slit selection was based on the seeing conditions and ranged from $1.1\arcsec$ to $5.1\arcsec \times 51\arcsec$.  This setup resulted in spectral resolutions of $\sim20$--$30$~{\AA} with useful wavelength coverage in the range $\lambda\lambda$4000--7600~{\AA}.  A standard Hoya L38 blocking filter was used to ensure that the first-order spectrum was not contaminated by second-order light for $\lambda \gtrsim 7600$~\AA. The efficiency of the waveplate as a function of wavelength was measured and corrected at all epochs by inserting a fully polarizing Nicol prism into the beam above the slit. The orientation of the slit on the sky was always set to a position angle of $0^{\circ}$, aligned along north-south. A series of four separate exposures that sample 16 orientations of the waveplate yields two independent, background-subtracted measures of each of the normalised linear Stokes parameters, $Q$ and $U$. On some nights several such polarization sequences of SN~2009ip were obtained and combined, with the weighting of the individual measurements based on photon statistics. 

The instrumental polarization of SPOL and the Kuiper, Bok and MMT telescopes is extremely small ($<0.1$\%), as measurements of unpolarized standard stars over the past two decades have consistently shown. Still, to confirm, we observed one or both of the unpolarized standard stars BD+28$^{\circ}$4211 and G191B2B (Schmidt et al. 1992b) during each epoch, confirming that the instrumental polarization has to be less than 0.1\% and that there is no low-level residual structure observed in the $Q$ and $U$ spectra.  The linear polarization position angle on the sky ($\theta\/$) was determined by observing one or more of the interstellar polarization standards Hiltner~960, VI~Cyg~\#12, or BD+59$^{\circ}$389 (Schmidt et al. 1992b) during all epochs. The adopted correction from the instrumental to the standard equatorial frame for $\theta\/$ for all epochs was determined from the average position angle offset of the polarized standards. Differences between the measured and expected polarization position angles were $<0\fdg3$ for all of the standard stars.

The Sep. 21 MMT observations were impacted by intermittent cloud cover, which degraded the signal-to-noise ratio of these observations. Nonetheless, secure detections of polarization from SN~2009ip were obtained, owing to the dual-beam design of SPOL, which allows for variations in atmospheric transparency during integrations. Measurements at all other epochs from Kuiper and Bok were obtained under clear conditions. From Arizona latitudes, SN~2009ip was observed through airmass values of 2.05--2.32. 

\subsection{The Kast spectrograph at Lick Observatory}
The Shane 3~m reflector at Lick Observatory equipped with the Kast spectrograph (Miller \& Stone 1993) was used on Oct. 5 and 14, and on Nov. 6 and 14. Like SPOL, Kast is a dual-beam spectropolarimeter that utilises a rotatable half-waveplate and Wollaston prism.  Only the red channel of Kast was used for spectropolarimetry. A GG455 order-blocking filter blocked all second-order light at wavelengths shortward of 9000~\AA. Observations were made with the 300~line~mm$^{-1}$ grating and the 3{\arcsec} slit, yielding a spectral resolution of $\sim$20~\AA.  The orientation of the slit on the sky was always set to a position angle of $0^{\circ}$, aligned along north-south. Exposures of 900~s were obtained at each of four waveplate positions ($0\fdg0$, $45\fdg0$, $22\fdg5$, and $67\fdg5$). For the Oct. 14, Nov. 6, and Nov. 14 observations, three sequences were performed yielding a total on-source exposure time of 3~hr. The observations on Oct. 5 were hampered by cloudy weather, and were halted after only one waveplate sequence (1~hr total exposure). From the latitude of Lick Observatory, SN~2009ip rose to a minimum airmass of 2.47 at transit. Observations typically began at airmass values near 2.5 and ended near 3.2.  Flat-field and wavelength calibration spectra were obtained immediately after each sequence, without moving the telescope. 

For polarimetric calibrations, standard stars were selected from the sample of Schmidt et al. (1992a,b). The unpolarized standard stars BD+32$^{\circ}$3739 and HD~57702 were observed to verify the low instrumental polarization of Kast. We constrained the average fractional Stokes $Q$ and $U$ values to $<0.05$\%. By observing BD+32$^{\circ}$3739 through a 100\% polarizing filter, we determined that the polarimetric response is so close to 100\% that no response correction was necessary, and we obtained the instrumental polarization position-angle curve, which we used to correct the data. We observed the high-polarization stars BD$+$59$^{\circ}$389 and HD~204827 to obtain the zero-point of the polarization position angle on the sky ($\theta$) and to determine the accuracy of polarimetric measurements, which were consistent with published values within 0.05\%. 

Since the Oct. 5 observations were hampered by intermittent cloud cover, no polarization standard stars could be observed. Thus, for this epoch we assumed negligible instrumental polarization, relying on the fact that the instrumental polarization was below 0.05\% for the Oct. 14, Nov. 6, and Nov. 14 observations, and we used the position-angle correction curve from Oct. 14,  which we have verified to be very stable over time. We also adopted an offset angle of 5.9$^{\circ}$ to further calibrate the Oct. 5 observations; this is the same offset value measured in the Oct. 14, Nov. 6, and Nov. 14 observations, to within 0.5$^{\circ}$. 
 
All data obtained from the Arizona and Lick observatories were extracted and calibrated using generic {\sc iraf} routines and our own {\sc idl} functions. Spectropolarimetric analysis was also performed in {\sc iraf} and {\sc idl} following the methods described by Miller et al. (1988) and implemented by Leonard et al. (2001). 

\subsection{The Very Large Telescope and FORS2}
We also obtained spectropolarimatric data on SN~2009ip from the European Southern Observatory (ESO) public archive. We analysed data from the Very Large Telescope (VLT) and Focal Reducer and low-dispersion Spectrograph (FORS2) instrument (Appenzeller et al. 1998) obtained on 2012 Nov. 7, Dec. 6, and Dec. 10, which were part of the target-of-opportunity programme ID 290.D-5006.  We used the 300V grism data for Nov. 7 and Dec. 6; these data have spectral resolution of $\sim$13~{\AA} and wavelength coverage of $\sim$4450--9500~\AA. We also analyzed higher-resolution ($\sim$3~{\AA}) 1200R grism data for dates Nov. 7 and Dec. 10, which covers a wavelength range of 5800--7300~\AA. The GG435 order-blocking filter was used for all observations.  Flat-field, bias frames, and wavelength-calibration spectra were obtained for both grism settings. A master distortion frame and a waveplate chromatism data file were also obtained from the archive.

SN~2009ip was observed at airmass values around 1.0--1.1 for the 300V and 1200R grism data, with respective exposure times of 330 and 470~s per waveplate position for the Nov. observations, and approximately 900~s for the Dec. observations. The position angle on the sky was always kept at 0$^{\circ}$ by the instrument rotator.  Standard stars from Fossati et al. (2007) were observed for calibrations. The high-polarization standard NGC~2024-1 was observed on Nov.~6 and Dec.~7 to calibrate the zero-point of the polarization position angle of the polarimeter on the sky, and the low-polarization standards WD~2039$-$202 and WD~2149$+$021 were observed on Nov.~1 and Dec.~7, respectively, to gauge any instrumental polarization. The instrumental position angle spectrum was obtained directly from the ESO public archive. On Nov.~1 the observation of WD~2039$-$202 yielded linear Stokes parameters of $(Q,U)<0.05$\%, after binning the $V$ band, while on Dec. 6, WD~2149$+$021 also yielded $(Q,U)<0.05$\%. Such a small amount of instrumental polarization will not have a substantial impact on the science observations, and we thus make no attempt to remove it.

Reduction and spectropolarimetric analysis of the VLT/FORS2 data was performed using the FORS pipeline kit (version 4.11.13) supplied by ESO. Specifically, the \textit{fors\_pmos\_calib} and \textit{fors\_pmos\_science} reduction recipes were used, and executed using the \textit{Gasgano} graphical user interface software. After all reduction and calibrations were performed, the data were corrected for the redshift of the host galaxy NGC~7259 ($z=0.005965$).
 
\section{Analysis \& Results}
Linear polarization is expressed as the quadratic sum of the $Q$ and $U$ Stokes parameters, $P= \sqrt{Q^2 + U^2}$, and the position angle on the sky is given by $\theta=1/2~\textrm{tan}^{-1}(U/Q)$, while carefully taking into account the quadrants in the $Q$--$U$ plane where the inverse tangent angle is located.  Since $P$ is a positive-definite quantity, it is significantly overestimated in situations where the signal-to-noise is low.  It is typical to express the ``de-biased" (or ``bias-corrected") form of $P$ as $P_{\rm db}=\sqrt{(Q^2 + U^2) - (\sigma_{Q}^2 + \sigma_{U}^2)}$, where $\sigma_Q$ and $\sigma_U$ are the uncertainties in the $Q$ and $U$ Stokes parameters.  If $(\sigma_{Q}^2 + \sigma_{U}^2) > (Q^2 + U^2)$, then we set a 1$\sigma$ upper limit on $P$ of  $\sqrt{\sigma_{Q}^2 + \sigma_{U}^2}$. In cases where $P/\sigma_{P} < 1.5$, $\theta$ is essentially undetermined.  All polarized spectra presented herein are displayed in this manner. Note, however, that at low signal-to-noise ratio $P_{\rm db}$ is also not a reliable function, as it has a peculiar probability distribution (Miller et al. 1988). Thus, for extracting statistically reliable values of polarization within a particular waveband, we have binned the calibrated $Q$ and $U$ Stokes spectra separately over the wavelength range of interest before calculating $P$ and $\theta$. All quoted and tabulated values in this paper were determined in this manner, although spectra are displayed as $P_{\rm db}$, so they may exhibit slight offsets from our quoted values.

\begin{figure}
\includegraphics[width=3.3in]{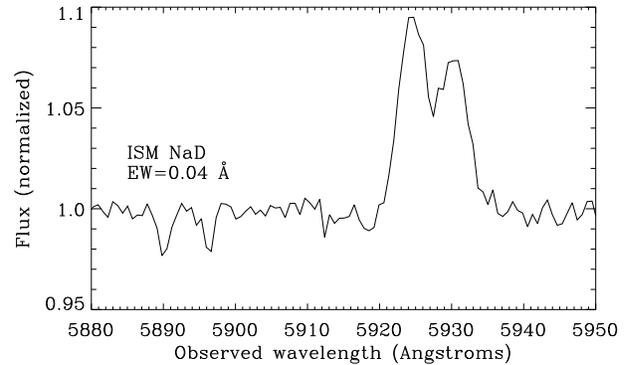}
\caption{VLT/FORS2 total-flux spectrum (normalised to the continuum) near the Na~D region, obtained on 2012 Nov. 7. The spectrum has not been corrected for cosmological redshift. Galactic interstellar Na~D absorption exhibits a very small equivalent width of $\sim$0.04~{\AA} for both components, implying low interstellar absorption and, hence, low Galactic interstellar polarization (see \S3.1). Na~D is seen in weak emission at the redshift of SN~2009ip.}
\label{fig:na}
\end{figure}

\subsection{Interstellar polarization}
Interstellar polarization (ISP) is commonly a very problematic element of polarimetric analysis, as an accurate probe of the ISP from both the Milky Way and the host galaxy can be very difficult to obtain. Fortunately, several factors indicate that the ISP in the direction of SN~2009ip must be very small, and should not significantly impact our polarimetry results and the physical interpretation that follows. 

First, the high-resolution VLT/FORS2 flux spectrum on Nov. 7, shown in Figure~\ref{fig:na}, reveals the resolved components of Galactic interstellar Na~I~D absorption lines at 5890.1~\AA\ and 5896.3~\AA. The equivalent width (EW) of each component is a very small $\sim40$~m{\AA}, which implies a Galactic $E(B-V)<0.01$~mag (Munari \& Zwitter 1997). According to Serkowski et al. (1975), the value of $E(B-V)$ can be used to derive an upper limit on Galactic ISP. The so-called Serkowski relation is given by ISP $<9E(B-V)$, which implies that ISP $<0.1$\% in the direction of SN~2009ip. 

Second, earlier photometric estimates of the total reddening toward the host galaxy NGC~7259 and SN~2009ip indicate a low value of $E(B-V)=0.019$ mag (Schlegel et al. 1998; Smith et al. 2010; Foley et al. 2011), which implies that the host galaxy  also does not produce substantial ISP along the line of sight. This is also consistent with the lack of Na~D absorption at the redshift of NGC~7259; in fact, Na~D appears in weak emission, perhaps from a local H~{\sc ii} region or from gas excited by SN radiation, with no indication of additional absorption components. 

The low host extinction is not particularly surprising, given SN~2009ip's large radial distance from the center of this nearly face-on spiral galaxy. If we were to naively apply the Serkowski relation to the total measured reddening, we would obtain ISP $<0.2$\%, still a very low value. However, as shown by Leonard et al. (2002), the Serkowski relation does not necessarily apply to all galaxies outside the Milky Way, as different dust properties can result in variable efficiencies for ISP. Nonetheless, the data strongly suggest a low value of total ISP, which justifies our decision to make no attempt to remove it from the data, as its effect should be minor and not significantly impact our results. We revisit the subject of ISP at the end of the following section.
 
 \begin{figure}
\includegraphics[width=3.3in]{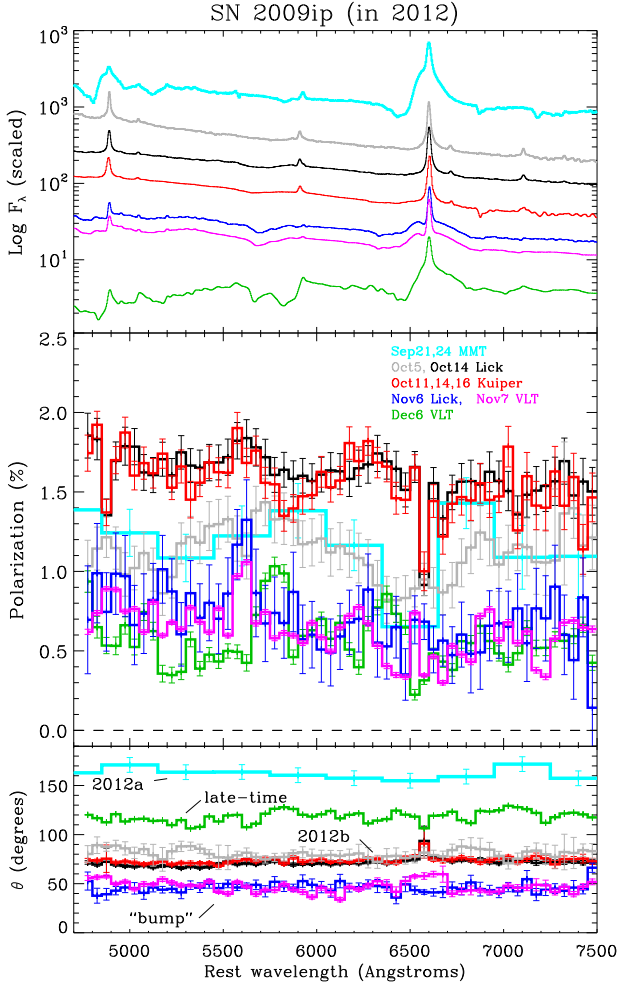}
\caption{Total flux, fractional debiased polarization $P_{\rm db}$, and position angle $\theta$ for SN~2009ip during its 2012 outburst, at epochs between Sep. 21 and Dec. 6. The three Kuiper/SPOL epochs (\textit{red}) have been averaged for clarity. The Sep. 21 and 24 MMT/SPOL data (\textit{light blue}) have been averaged and binned to 300~{\AA}; all other epochs are binned to 50~{\AA}. Note the $\sim90^{\circ}$ offset in $\theta$ between the 2012a phase and the peak of 2012b.}
\label{fig:pol_seq}
\end{figure}

\begin{figure}
\includegraphics[width=3.in]{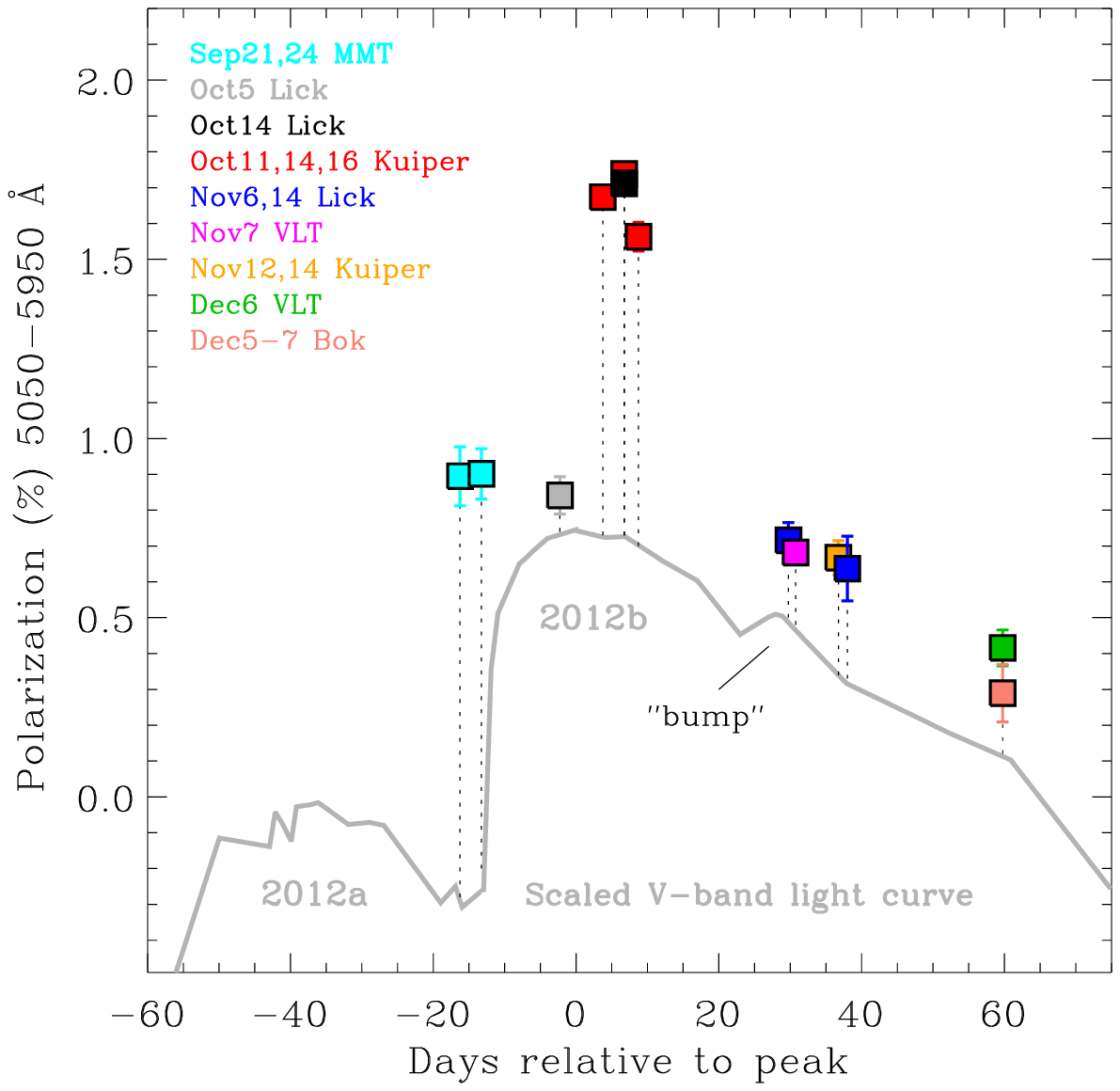}
\includegraphics[width=3.in]{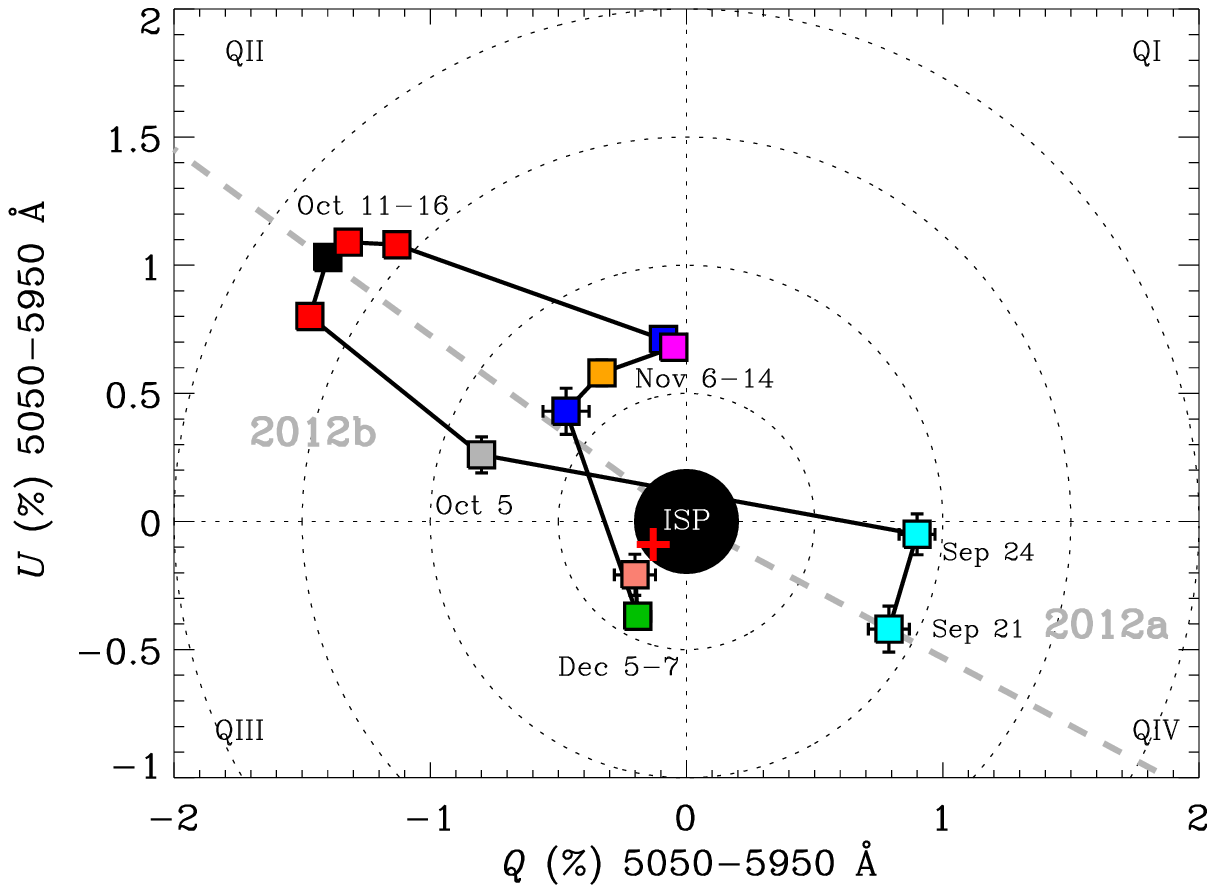}
\caption{\textit{Upper panel}: Temporal evolution of the total $V$-band (5050--5950~{\AA}) polarization for SN~2009ip during the 2012a and 2012b phases. For reference, an arbitrarily scaled version of the $V$-band light curve is plotted in the background. \textit{Lower panel}: Evolution in the $Q$--$U$ plane. Interstellar polarization is constrained to $<$0.2\%, illustrated by the black dot near the origin. The dashed lines represent the position angles $\theta_V=166^{\circ}$ and $72^{\circ}$ measured at the first epoch on Sep. 21 and peak polarization on Oct. 14, respectively. Our latest continuum measurement (5200--5500~{\AA}) for Dec. 6 is marked with a red cross near the origin, consistent with our ISP limit. The approximate point-reflection symmetry between 2012a and the peak of 2012b suggests two separate and roughly orthogonal components of polarization on the sky. Note the temporary shift in $\theta$ associated with the bump in the light curve at $\sim$30 days past peak.}
\label{fig:pol_lc}
\end{figure}

 \subsection{Spectropolarimetric evolution of SN 2009ip}

Figure~\ref{fig:pol_seq} shows the total flux spectra, $P$, and $\theta$ of SN~2009ip for epochs between Sep. 21 and Dec. 6.  To gauge the total $V$-band polarization $P$ and $\theta$, we first binned $Q$ and $U$ separately in the wavelength range 5050--5950~{\AA}.  The extracted values are listed in Table~1. Figure~\ref{fig:pol_lc} shows the resulting polarization light curve and the temporal evolution in the $Q$--$U$ plane. In addition to $V$-band, we also sampled ``continuum" regions of 5200--5500~{\AA} and 6100--6200~{\AA}, because they appear mostly devoid of strong line features; these ranges are close to the respective ``green" and ``red" wavelength regions used by Hoffman et al. (2008). We caution, however, that there is no location within the spectrum that is completely free of line emission/absorption during all epochs, so the regions we have selected should be regarded as pseudocontinua that avoid the strong lines.  

\begin{table*}\begin{center} 
      \caption{Spectropolarimetry of SN~2009ip}
\renewcommand\tabcolsep{14pt}
\begin{tabular}[b]{@{}lrccccr}
\hline
\hline
Date (2012 UT) &day$^{\textrm{a}}$ &Tel./Instr.   & $P_{\rm green}$(\%) &   $P_{\rm red}$(\%) &  $P_{V}$(\%) & $\theta_{V}$ (deg)  \\
 \hline
  \hline
Sep. 21             & $-13$         &MMT/SPOL         & 0.96 (0.10)  &      :::              & 0.89 (0.08)   & 166 (5)    \\
Sep. 24             & $-10$         &MMT/SPOL         & 1.15 (0.09)  & 0.99 (0.11)   & 0.90 (0.08)   & 178 (5)    \\
Oct. 5$^{\textrm{b}}$& $-3$ & Lick/Kast             & 0.81 (0.07)  &  1.03 (0.10)   & 0.84 (0.12)   & 81 (4)   \\
Oct. 11              &  4                &Kuiper/SPOL      & 1.63 (0.05)   &  1.66 (0.07)  &  1.67 (0.06)   & 76 (1)    \\
Oct. 14               &  7               &Kuiper/SPOL      & 1.82 (0.04)   &  1.57 (0.06)  &  1.71 (0.05)   & 71 (1)    \\
Oct. 14               &  7               &Lick/Kast              & 1.75 (0.05)  &  1.65 (0.05)   &  1.73 (0.05)   &  72 (1)   \\
Oct. 16             &  9                 &Kuiper/SPOL       & 1.59 (0.04)   &  1.75 (0.06)  &  1.56 (0.06)   & 68 (1)    \\
Nov. 6                &  30             &Lick/Kast              &  0.76 (0.04)  &  0.71 (0.07)  & 0.72 (0.06)   & 49 (3) \\
Nov. 7                &  31             & VLT/FORS2        &   0.67 (0.04)  &  0.70 (0.05) &  0.69 (0.03) &  47 (2)  \\
Nov. 12--14       &  36, 38     &Kuiper/SPOL&  0.61 (0.07) & 0.89 (0.05)  &  0.66 (0.06)  &   60 (5) \\
Nov. 14             &  38             &Lick/Kast               &  0.52 (0.13)  &    ::::               &  0.64 (0.09) &   66 (6)   \\ 
Dec. 6                 &  60             &VLT/FORS2         &  0.16 (0.04)  &  0.60 (0.05) &  0.41 (0.04) &  122 (2) \\
Dec. 5--7             & 59--61      &Bok/SPOL           &  $<$0.22       & 0.81 (0.13)  & 0.29 (0.10)   &  113 (6)  \\
Dec. 10 & 64      &VLT/FORS2                    &  :::                      & 0.56 (0.06)  &     :::                &  $\theta_{\rm red}=132$ (3)  \\
\hline
\end{tabular}\label{tab:p48} 
\end{center}
\begin{flushleft}
$^\textrm{a}$Day is with respect to the day of peak in the $V$ band (JD~2456207.72). \\
$^\textrm{b}$Data for Oct. 5 interrupted by poor weather; required calibration using Oct. 14 standard-star observations; to be used with caution.   \\
\end{flushleft}
\end{table*}

The Sep. 21 and 24 epochs took place near the end of SN~2009ip's 2012a phase, before the sharp rise in brightness that marks the onset of the 2012b phase. The total flux spectrum during 2012a exhibits the characteristics of SNe~IIn, displaying intermediate-width ($\sim$1000~km~s$^{-1}$) emission lines superimposed on a spectrum with broad P-Cygni lines ($\sim$8000~km~s$^{-1}$) that is reminiscent of more common SNe~II-P (Mauerhan et al. 2013). During this early epoch, SN~2009ip is polarized at $\sim$0.9\% with position angle $\theta\approx166^{\circ}$, and it occupies quadrant~{\sc IV} in the $Q$--$U$ plane. 

By Oct. 5, SN~2009ip has entered the 2012b phase and the flux spectrum changes substantially; the broad emission components have become strongly diluted by the strengthening continuum, which is quite blue, while the intermediate-width emission features have developed broad Lorentzian wings, a commonly observed characteristic of SNe IIn and an indication of high optical depth (Chugai 2001; Dessart et al. 2009; Smith et al. 2008), probably resulting from CSM interaction. At this time, $P$ has changed only marginally to $\sim$0.8\% but  the position angle has shifted dramatically to $\theta\approx81^{\circ}$, moving into quadrant~{\sc II} of the $Q$--$U$ plane. Interestingly, this dramatic change in position angle indicates separate polarization components for the 2012a and 2012b phases, with distinct axes of symmetry. The lack of a correspondingly significant change in fractional $P$ associated with the large rise in brightness could possibly be explained by the partial cancellation of the polarization vectors from the 2012a and 2012b components, as the latter rises in luminosity. 

By Oct. 11, the source has risen to peak flux. The spectrum has retained the overall appearance of Oct. 5, while the continuum becomes less blue and broad-line P-Cygni absorption becomes apparent for He~{\sc i}/Na~D. Our Lick/Kast and Kuiper/SPOL measurements on Oct. 14  show that $P$ has risen to a maximum of $\sim$1.7\% and $\theta$ has reached $\sim72^{\circ}$. At this time, strong deficits in polarization are seen for the intermediate-width emission components of H$\alpha$ and, to a lesser extent, H$\beta$. This is likely to be the result of dilution from the strong, intrinsically unpolarized emission lines in the outer CSM. Meanwhile, a slight increase in $P$ is seen at 5600--5800~\AA, suggesting a potential association with broad He~{\sc i} $\lambda$5876 or Na~D P-Cygni lines. 

\begin{figure*}
\includegraphics[width=6.5in]{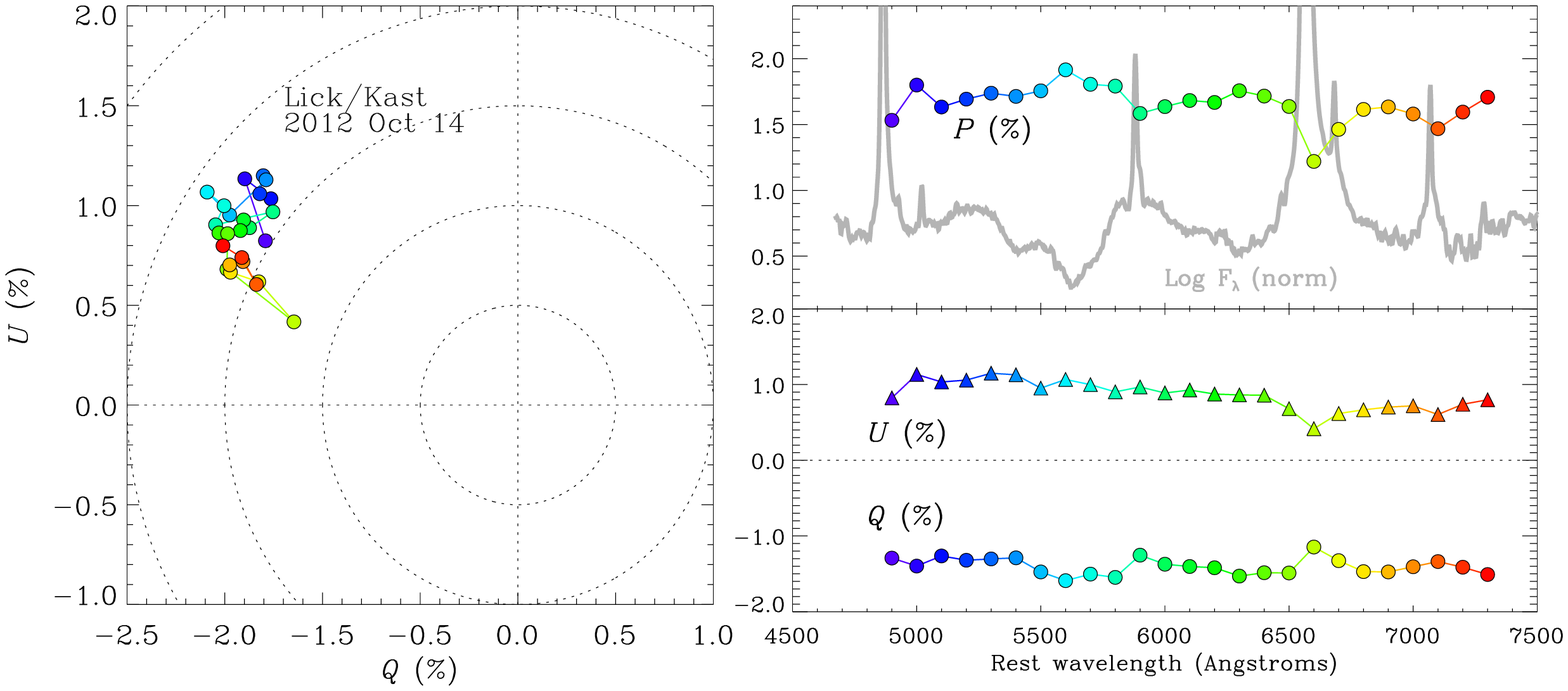}
\includegraphics[width=6.5in]{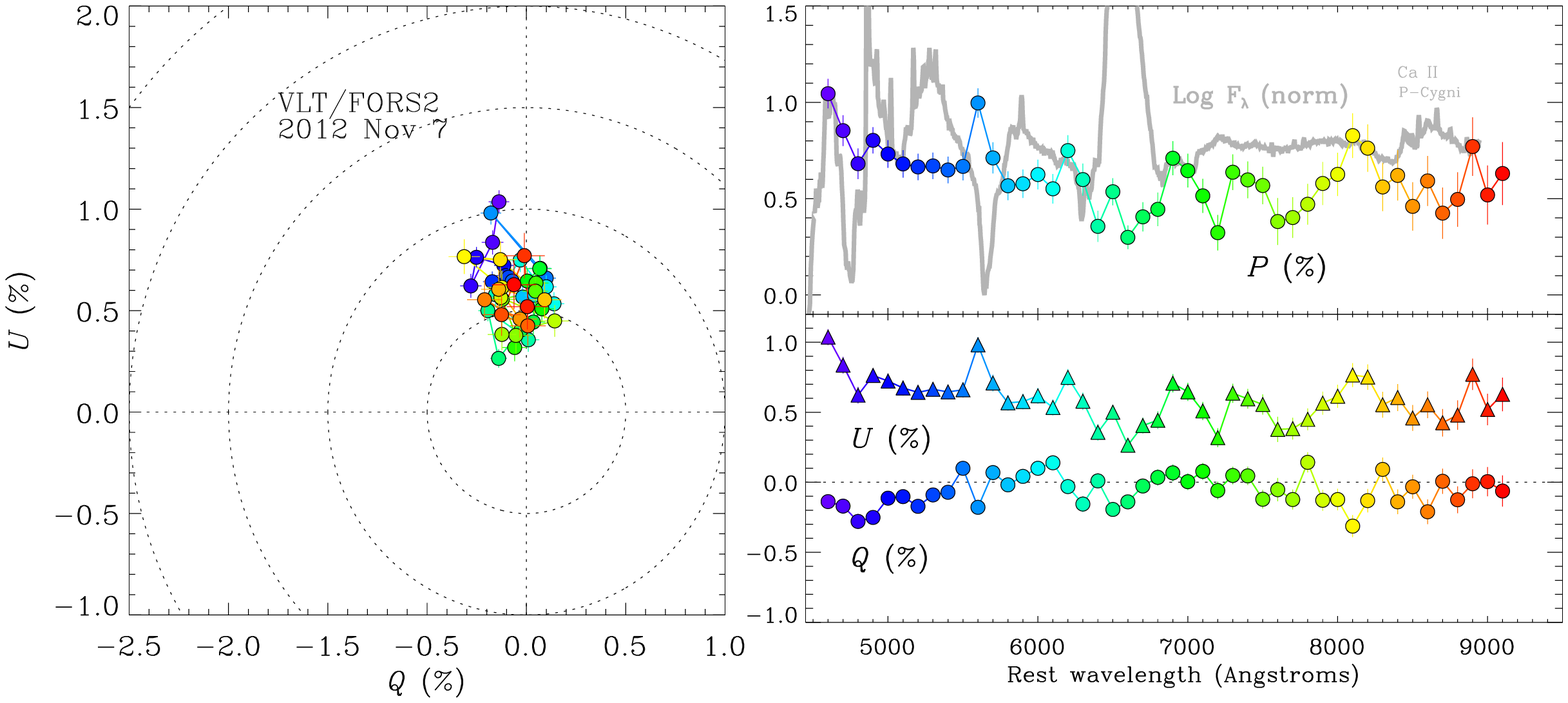}
\includegraphics[width=6.5in]{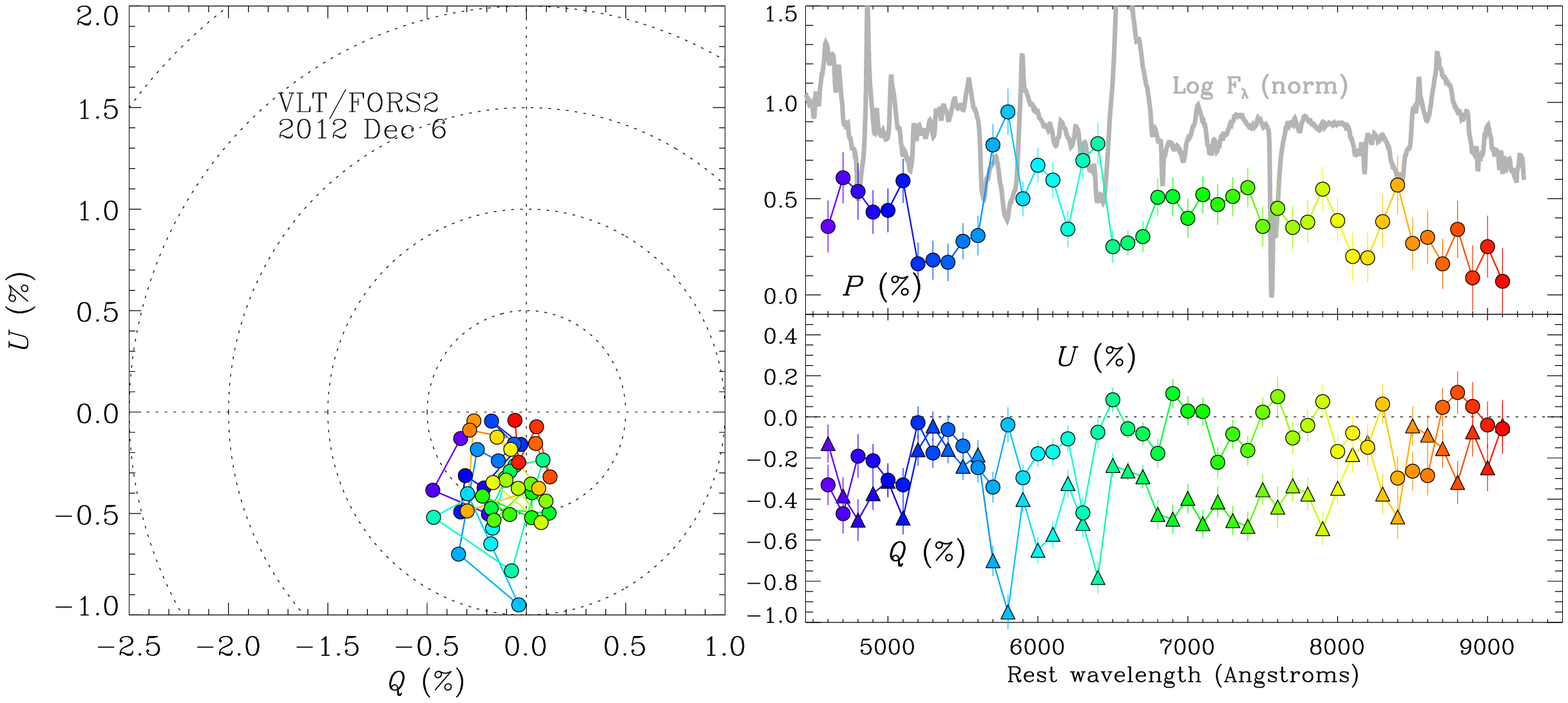}
\caption{$Q$ and $U$ Stokes parameters and polarization for SN~2009ip on 2012 Oct. 14 (from Lick/Kast, upper panels), Nov. 7 and Dec. 6 (VLT/FORS2, 300V grism, middle and lower panels). The colors have been chosen to correspond with wavelength, but note the different wavelength scales for the Lick and VLT data. The data have been binned to 100~{\AA}. Scaled versions of the total flux spectra are plotted in the background for reference (solid grey curves).}
\label{fig:qu_full}
\end{figure*}

Figure~\ref{fig:qu_full} (top panels) shows the Stokes spectra and the data distribution in the $Q$--$U$ plane for the Oct. 14 Lick epoch, binned to 100~{\AA}. For reference, we also show scaled logarithmic flux spectra in the same figure, which has been normalised with a low-order spline (avoiding the strongest emission features for the spline nodes). On Oct. 14, near peak brightness, the entire optical spectrum of SN~2009ip occupies a compact portion of the $Q$--$U$ plane, in quadrant~{\sc II}. The points form a slightly elongated cluster that shows a continuous trend in wavelength, with the bluest points exhibiting the lowest position angle in the $Q$--$U$ plane. The cluster as a whole exhibits a small but non-negligible spread in position angle, which corresponds to a range of $\sim10^{\circ}$ on the sky. The dilution in polarization associated with intermediate-width H$\alpha$ emission is apparent as a protrusion from the cluster of data points which roughly points toward the origin. Note that our binning causes the line to be unresolved, and this affects the strength and position angle of the protrusion; the unbinned data show the protrusion extending very close to the origin.  In total polarization $P$, we see broad, shallow variations that appear to be associated with broad absorption components of H$\alpha$ and He~{\sc i}/Na~D blueward of 5876~\AA. It is interesting to note that the scaled flux spectrum included in Figure~\ref{fig:qu_full} shows that the broad P-Cygni emission features from fast ejecta, first seen during 2012a, are still present throughout the peak of 2012b; this has important physical consequences for the energetics of SN~2009ip (see \S4).  

\begin{figure*}
\includegraphics[width=5.5in]{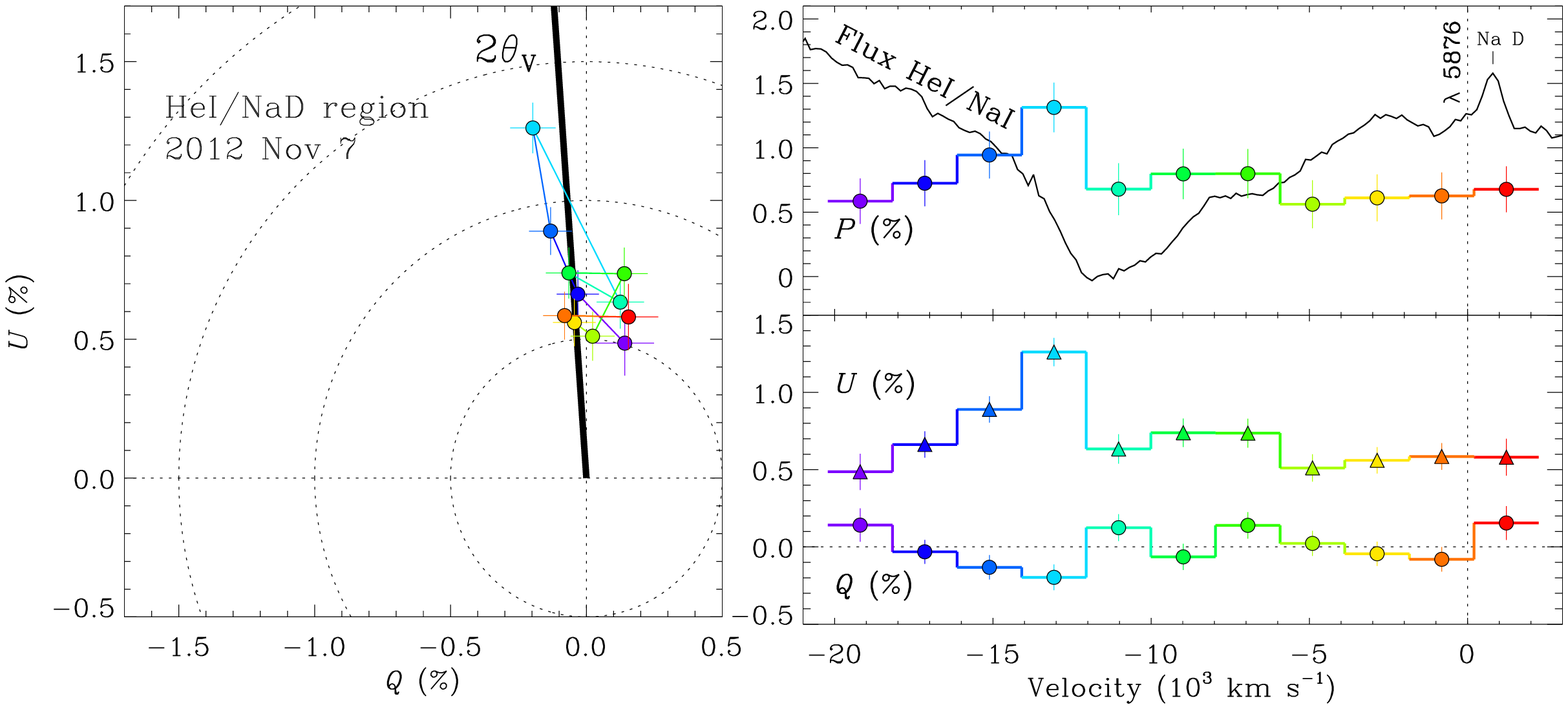}
\includegraphics[width=5.5in]{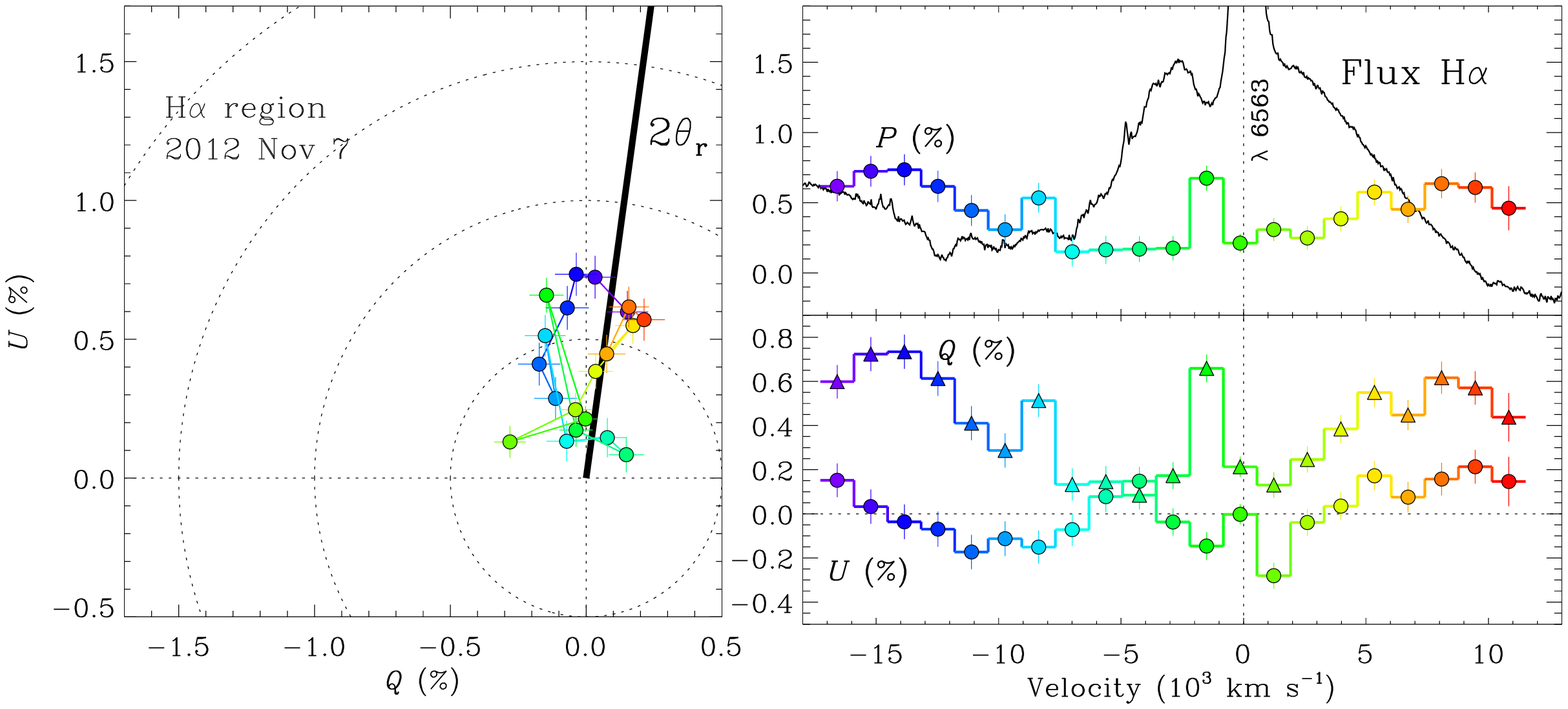}                  
\caption{VLT/FORS2 $Q$ and $U$ Stokes parameters and polarization for SN~2009ip on 2012 Nov. 7 for spectral regions near 5876~\AA\ (upper panels; 300V grism; binned to 40~{\AA}) and H$\alpha$ (lower panels; 1200R grism; binned to 30~{\AA}). The right-hand plots are with respect to velocity. The colors have been chosen to correspond with velocity, but note the different velocity scales for the upper and lower panels. Scaled versions of the total-flux spectra are plotted in the background for reference (black curves in right-hand panels). Enhanced polarization is associated with He~{\sc i} $\lambda$5876 (or Na~D) P-Cygni absorption, at a blueshifted velocity of $-$13,000~km~s$^{-1}$. The enhancement forms a linear loop in the $Q$--$U$ plane that roughly follows along the $V$-band (5050--5950~\AA) symmetry axis for that epoch ($2\theta_{V}$, thick black line). Note that the weak emission line redward of 5876~\AA\ is Na~D emission (see Figure~\ref{fig:na}). For H$\alpha$, enhanced polarization appears to be associated with the intermediate-width absorption feature of H$\alpha$, while the broad underlying feature exhibits a loop in the $Q$--$U$ plane, within the range of position angles bounded by $2\theta_r$ and $2\theta_V$. }
\label{fig:qu_line1}
\end{figure*}

\begin{figure*}
\includegraphics[width=5.5in]{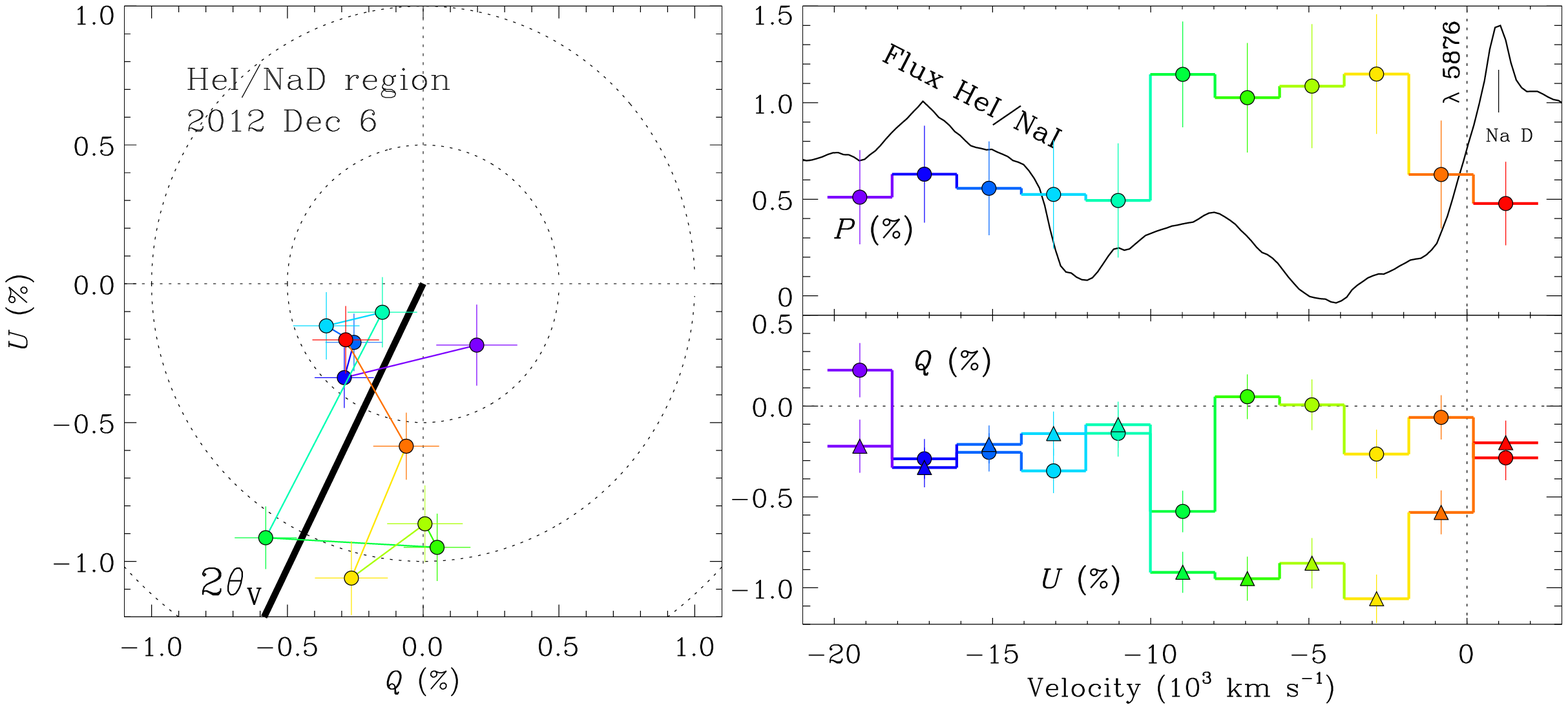}
\includegraphics[width=5.5in]{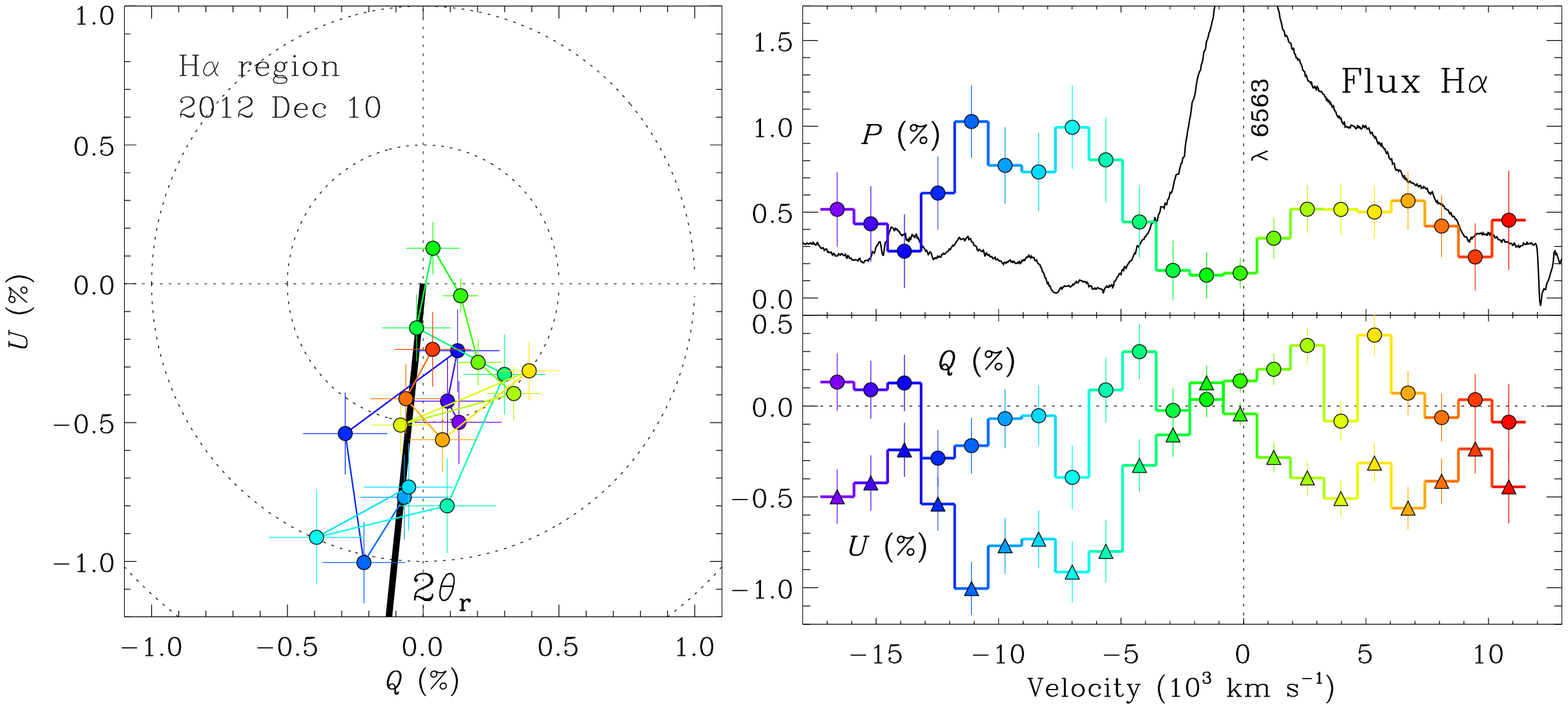}                  
\caption{Same a Figure~\ref{fig:qu_line1} but for later epochs. On Dec. 6--10 (lower two panels), enhanced polarization is now associated with the lower velocity component of the double-dipped absorption feature, produced by either He~{\sc i} or Na~D. Enhanced polarization is also seen for the blend of absorption components blueward of the H$\alpha$ line center. The He~{\sc i}/Na~D and H$\alpha$ features both appear to make loops in the $Q$--$U$ plane, roughly along the average symmetry axes of the given epochs.}
\label{fig:qu_line2}
\end{figure*}

By Nov. 6--7, roughly 30 days past peak, the continuum has faded substantially and broad P-Cygni lines have once again become prominent in the flux spectrum. He~{\sc i}/Na~D absorption has deepened, and the intermediate-width component of H$\alpha$ has developed an absorption component on top of the broad underlying feature, perhaps part of its own P-Cygni profile. Interestingly, this epoch coincides with the appearance of a temporary albeit substantial bump in the declining light curve. The Nov. 7 VLT/FORS2 spectropolarimetry data are shown in detail in the lower panels of Figure~\ref{fig:qu_full}. At this time, the $V$-band polarization has dropped to $\sim$0.7\% and the position angle has shifted substantially to $\sim$50$^{\circ}$. The enhancement of polarization near He~{\sc i}/Na~{\sc i} absorption feature has become more pronounced and sharp; this feature is also apparent in Lick data from Nov. 6, shown in Figure~\ref{fig:pol_seq}. Meanwhile, the sharp dip in polarization produced by the dilution from unpolarized intermediate-width Balmer lines is no longer apparent.

We examine the polarization enhancements of two specific line features in Figure~\ref{fig:qu_line1}, which show the Stokes parameters plotted as a function of velocity, with respect to the He~{\sc i} $\lambda$5876 and H$\alpha$ rest wavelengths. On Nov. 7, the enhanced polarization of He~{\sc i}/Na~D is substantially blueshifted to a velocity of 13,000~km~s$^{-1}$ (if it is associated with He) and exhibits a velocity width of $\sim$1500~km~s$^{-1}$. The peak is also blueshifted by $\sim$1500~km~s$^{-1}$ with respect to the flux minimum of the absorption line. The enhanced He~{\sc i}/Na~D feature forms a linear-shaped loop that roughly traces the average polarization axis of the system for that date. 

There also appears to be an enhancement in fractional polarization for the absorption component blueward of the  intermediate-width H$\alpha$ emission feature (at $-1500$~km~s$^{-1}$). The origin of this absorption feature is unclear. It could be an absorption component of a P-Cygni profile for the intermediate-width H$\alpha$ component, or it could be the result of blueshifted absorption for He~{\sc i}~$\lambda$6678. If it is from H$\alpha$ P-Cygni, then this enhanced polarization feature does not appear blueshifted as in the case of the broad He~{\sc i}/Na~D absorption. In the $Q$--$U$ plane, the track across this potential H$\alpha$/He~{\sc i} feature appears to be roughly consistent with the overall axis of symmetry. There also appears to be some indication of polarization enhancement for blend of absorption components seen at the higher velocities of $\sim10^4$~km~s$^{-1}$ and larger, which are probably from H$\alpha$. Overall, the broad emission component of H$\alpha$ appears to make a loop in the $Q$--$U$ plane.

Averaging all of our measurements from Kuiper/SPOL on Nov. 12 and 14, we detect {$\sim$0.7\%} polarization in the $V$ band, similar to the value of the preceding VLT/FORS2 epoch on Nov. 7. The Lick/Kast data solely from Nov. 14 show a value of polarization similar to the Nov.12--14 combined average of the Kuiper/SPOL data. The polarized spectra are not shown in Figure~\ref{fig:pol_seq}, to avoid cluttering the figure, although their broadband measurements are presented in Figure~\ref{fig:pol_lc}. Interestingly, although the overall degree of polarization has not changed substantially  between Nov. 6 and 14, the position angle has shifted again back toward the axis of symmetry of the 2012b peak. It thus appears that the bump in the light curve on Nov. 6--7 is associated with another component of polarization, perhaps related the 2012a luminosity source.

A few weeks later, on Dec. 5--7, we detect weak polarization in an average of three separate measurements with Bok/SPOL, obtaining $P_V\approx0.3\%$ and only an upper limit for the continuum of $P_{\rm green}<0.22$\%. The last VLT/FORS2 measurement takes place on Dec. 6, for which we obtain $P_{\rm green}\approx0.16$\% and $P_V\approx0.4$\%. For the December epochs the position angle has changed significantly with the diminishing polarization, moving into quadrant~{\sc III} with $\theta\approx120^{\circ}$. The flux spectrum during this late epoch shows that the continuum has become redder, while the broad P-Cygni features are still present. He~{\sc i}/Na~D absorption has become a double-dipped profile. Figure~\ref{fig:qu_line2} shows that by Dec. 6 the enhanced polarization of this feature has shifted redward, and is probably associated with the bluer component of the double-dipped profile. A complicated blend of multiple absorption and perhaps emission features appears blueward of H$\alpha$ at this time, and also exhibits evidence for enhanced polarization, roughly along the average axis of symmetry. 

\subsubsection{ISP revisited}

Examining more closely the temporal evolution of SN~2009ip in the $Q$--$U$ plane, shown in Figure~\ref{fig:pol_lc} (lower panel), the range of possible ISP values constrained by the value of $E(B-V)$ (see \S3.1) are illustrated by the radius of the black spot centered on the origin. It seems plausible that the path traced by $Q$ and $U$ during Dec. is the result of a migration back to the point of ISP, somewhere near the origin. 
Indeed, our latest measurement of the continuum polarization yielded $P_{\rm green}<0.22$\% from Bok/SPOL for Dec. 5--7, and $P_{\rm green}=0.16$\% from VLT/FORS2 on Dec. 6 ($q=-0.13$\%, $u=-0.09\%$), which is consistent with the ISP limits we have estimated. An alternative, and less likely possibility, is that $Q$ and $U$ are destined to drift back to the location in quadrant~{\sc IV} occupied by our earliest MMT/SPOL measurements in September (i.e., that the 2012a phase is intrinsically unpolarized, and instead reflects the location of ISP). But such a large value of ISP ($\gtrsim$1\%) seems implausible given the constraints on ISP indicated by the low values of $E(B-V)$ and EW(Na~D); the structure in $P$ for the September epochs also does not  follow a Serkowski law, which is inconsistent with the 2012a polarization being the result of ISP.  Furthermore, it would be a rather remarkable coincidence for the position angle of ISP to be orthogonal to the polarization axis of SN~2009ip at peak. 

Therefore, the evidence is strong that our earliest polarization detections from MMT/SPOL in September reveal true polarization intrinsic to the source during the 2012a phase, and that the ISP is within $<0.2$\% of the origin, and likely $\lesssim0.16$\%, as indicated by our latest continuum measurement.  As one can see, any value of ISP within this allowable range will affect our results only slightly, and will not change the fact that the 2012a and 2012b phases appear to have two separate components of polarization that will remain roughly orthogonal between 2012a and the peak of 2012b, no matter where the ISP lies within the range illustrated by Figure~\ref{fig:pol_lc} (lower panel). 
 
 \begin{figure*}
\includegraphics[width=7in]{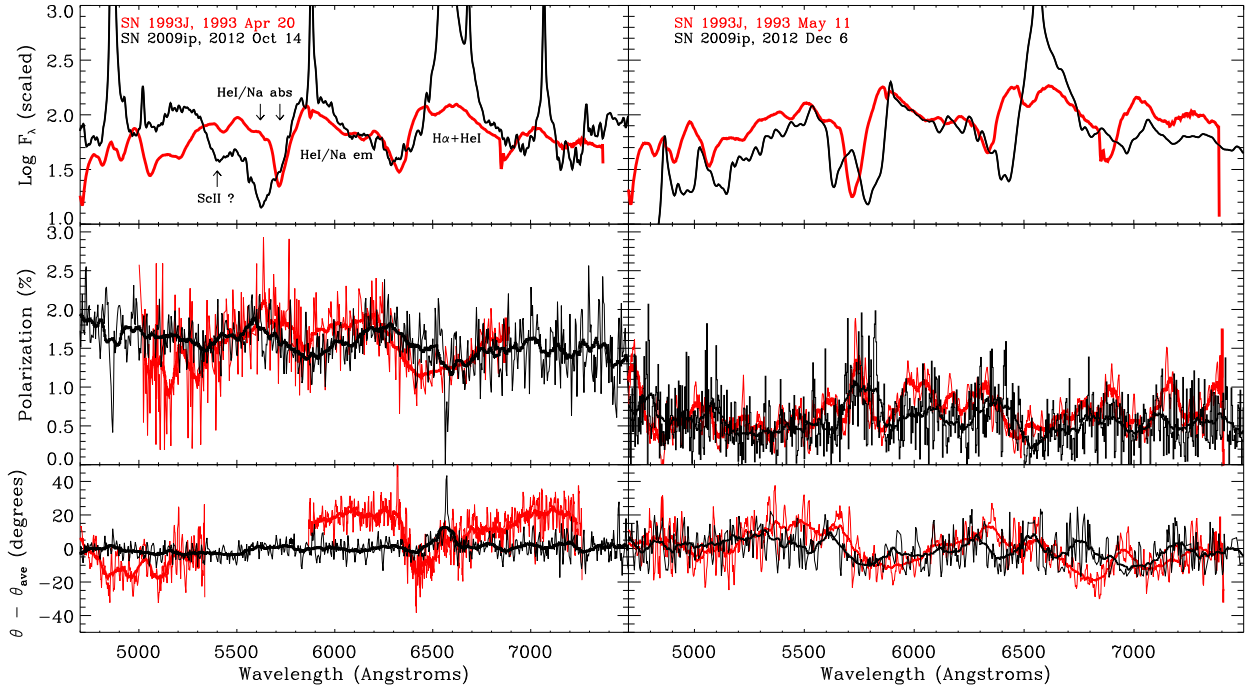}
\caption{Comparison between the flux and polarization properties of SN~2009ip (\textit{black}) and SN~1993J (\textit{red}). Both unbinned and smoothed versions of the data are shown for each SN. No scaling or adjustments have been applied. The SN~1993J April~30 data were traced from H{\"o}flich et al. 1996) and the May 11 data are from Tran et al. (1997). For the earlier epochs, which are near the main photometric peaks of both SNe, they appear very similar in spectral polarization. SN~2009ip exhibits little deviation in $\theta$ while the variations for SN~1993J (Tran et al. 1997) are more pronounced. At 50--60 days past peak polarization, both SNe exhibit very similar variations across their spectra, and their position angle variations become more comparable.}
\label{fig:93j}
\end{figure*}

\section{Discussion}
\subsection{Sources of polarization}
The most natural explanation for linear polarization of a SN is Thomson scattering by free electrons, either from the SN photosphere or within the zone of intense CSM interaction, which includes the forward shock, the reverse shock, and the cold dense shell (CDS) in between. For most SNe~IIn, including SN~2009ip, evidence of electron scattering is also indicated by the broad Lorentzian wings typically seen at the bases of the intermediate-width emission-line profiles (Chugai et al. 2001; Smith et al. 2008; Dessart et al. 2009). But in the case of scattered photons emerging from a spherically symmetric distribution of material, the sum of all electric vectors will cancel out and yield zero net polarization. Thus, the detection of net polarization from any SN directly implies aspherical structure, either intrinsic or imposed from partial obscuration of the polarized surface by intervening material (see Wang \& Wheeler 2008 for a review on SN polarization).

Core-collapse SNe of Type II-P have been observed to be weak sources of polarization while in their plateau phases, which implies that the outer layer of ejecta is roughly spherical in some cases. However, near the end of their plateaus, SNe II-P can exhibit brief but significant increases in polarization (Leonard et al. 2000, 2006; Chornock et al. 2010), because as the ejecta cool and recombine the photosphere recedes from the spherical outer layers of H-rich material into more aspherical He-rich and metal-rich layers deeper inside the ejecta. Thus, to first order, SNe II-P can be characterized by a single component of photospheric polarization that scales inversely with the decreasing luminosity. SNe~IIn, on the other hand,  appear to exhibit a high degree of polarization while in their brightest phases, at least for the few objects that have been studied in sufficient detail (e.g., Leonard et al. 2001; Hoffman et al. 2008). This implies that CSM interaction luminosity must be a major source of the net polarization from SNe~IIn. In this case, since both the SN photosphere and the CSM interaction zone are potential sources of polarized photons, changes in the relative luminosities of these components can lead to variations in net polarization, even if the fractional polarization and position angles of each component remain constant with time. Below, we propose this to be the case for SN~2009ip.

\subsubsection{Comparison to SN 1993J: evidence for a photospheric component}
SN~2009ip's photospheric component is made evident by its spectropolarimetric similarities with the Type IIb SN~1993J, shown in Figure~\ref{fig:93j}. Near peak brightness, both SNe exhibit a comparable degree of polarization across the optical band, with SN~2009ip being  stronger in the continuum, and they both display remarkably similar wavelength-dependent variations across their polarization spectra, particularly evident at wavelengths of 5200--7000~{\AA}. In the case of SN~1993J, these variations are associated with the broad P-Cygni lines of H$\alpha$ and He~{\sc i}$+$Na~D, and the same is probably true for SN~2009ip. The variations are produced by the depolarizing effect of line scattering (Trammell et al. 1993; H{\"o}flich 1995; H{\"o}flich et al. 1996; Tran et al. 1997). This is not to be confused with the intrinsically unpolarized recombination emission from the outer optically thin CSM that we see in SN~2009ip, which dilutes the polarized continuum and results in relatively narrow ``absorption" features of polarization for the H$\alpha$ and H$\beta$ lines. Rather, true depolarization is important for the broad emission lines that develop in the ionised ejecta above the electron-scattering photosphere of the SN. The persistence of these lines in SN~2009ip for $\sim$100 days implies a massive amount of high-velocity material (a few M$_{\odot}$), consistent with a SN outflow (see SMP14).

CSM is also a potential contributor to the polarization properties of SN~1993J (H{\"o}flich et al. 1996; Tran et al. 1997), via electron scattering within a SN/CSM interaction zone and/or the scattering of photons by dust. After all, the continuum polarization of SN~1993J being entirely due to an aspherical ejecta photosphere is somewhat difficult to reconcile with the evolution of a highly spherical radio remnant revealed by VLBI (Bietenholz, Bartel, \& Rupen 2003), although this structure might only trace the morphology of the outermost H-rich layers, not the more asymmetric He-rich layers potentially responsible for the polarized continuum. Still, the source did exhibit evidence for pre-existing CSM at early times, in the form of narrow emission-line features exhibiting small velocity widths of $\sim$170~km~s$^{-1}$, indicative of an ambient stellar wind (Benetti et al. 1994). However, these features were less pronounced than those of SN~2009ip and lasted for a shorter amount of time, indicating a much less dense CSM.  SN~1993J's X-ray properties were also indicative of CSM interaction (Leising et al 1994), while its radio evolution indicated a changing pre-SN mass-loss rate in the range $\dot{M}=10^{-5}$--$10^{-4}~{\rm M}_{\odot}$~yr$^{-1}$ (Van Dyk et al. 1994), which, although roughly 100 times weaker than the pre-SN mass-loss rate inferred for SN~2009ip (Ofek et al. 2013b), could generate enough CSM to influence the object's polarization properties. 

Therefore, like SN~2009ip, SN~1993J could have two components of polarization --- photospheric and  CSM interaction. In the case of SN~2009ip, however, it appears rather obvious that CSM interaction is the dominant source of polarization near photometric peak. The opposite could be true for SN~1993J: its polarization could be dominated by an aspherical explosion and only moderately influenced by the CSM component. In this regard, a noteworthy difference between the two SNe is that near peak brightness SN~1993J exhibits more variation than SN~2009ip in $\theta$ as a function of wavelength, suggesting that SN~2009ip has a more highly ordered axis of symmetry in 2012b resulting from its luminous and highly aspherical CSM interaction zone. By 50--60 days after peak, however, both SNe exhibit remarkable similarity in $P$ and $\theta$ as a function of wavelength. It appears that as CSM interaction fades after the peak of 2012b and the broad P-Cygni lines return by Nov. 6, the polarization characteristics of SN~2009ip begin to look more like a SN photosphere again. It is thus interesting that the source makes a pronounced shift in position angle on Nov. 6, associated with the obvious bump in the light curve (see Figure~\ref{fig:pol_lc}), which could represent the temporary brightening of the photospheric component of polarization that dominated the 2012a phase. 

\subsubsection{Orthogonal geometric components}
Taken at face value, SN~2009ip's degree of peak polarization for the 2012a and 2012b phases (from $P\approx0.9$\% to $1.7$\%) indicates respective axial ratios of $<$0.85 and $<$0.7 for the aspherical source ($>$15\% and $>$30\% asphericity), if we assume that polarization translates directly to the apparent geometry via comparison with the oblate electron-scattering atmosphere models of H{\"o}lfich (1995). However, the true physical axial ratios could be substantially more aspherical than these limits, for two key reasons: (1) inclination angle, and (2) partial cancellation between the 2012a and 2012b polarization vector components. If both the 2012a peak and the late decline of the 2012b phase (from the Nov. 6--7 bump onward) are dominated by the SN photosphere component, as appears to be the case, then partial cancellation could be particularly important for the 2012b peak. Indeed, the $\sim90^{\circ}$ shift in position angle between the earliest 2012a observation and the peak of the 2012b phase implies two separate components of polarization that are roughly orthogonal to one another. Since orthogonal polarization vectors cancel, the degree of intrinsic polarization for the 2012b component must be higher than it appears (i.e., even more aspherical), because the orthogonality of the separate 2012a vector component, if still present during 2012b, will partially cancel the 2012b component. 

\begin{figure}
\includegraphics[width=3.3in]{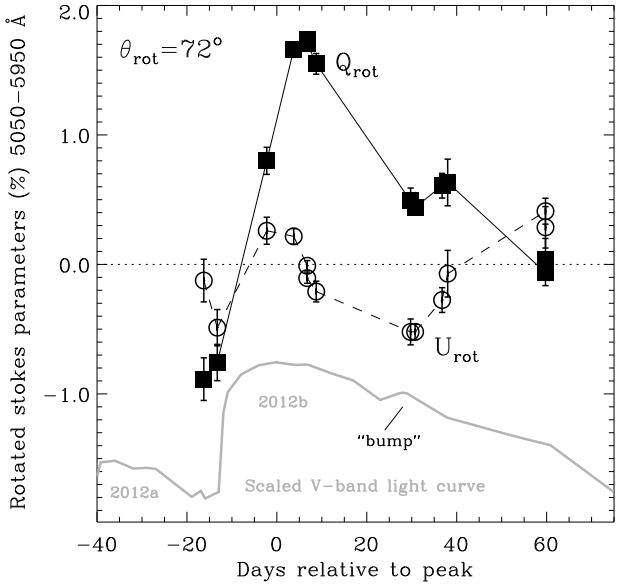}
\caption{Temporal evolution of the rotated Stokes parameters, aligned to the axis of symmetry exhibited during maximum polarization in 2012b phase ($\theta=72^{\circ}$). The plot is meant to illustrate the full possible degree of intrinsic polarization for the 2012b component ($\sim$2.6\%), after considering the effect of the partial cancellation, roughly orthogonal 2012a vector component (see \S4.1.2).}
\label{fig:rsp}
\end{figure}

\begin{figure}
\includegraphics[width=3.3in]{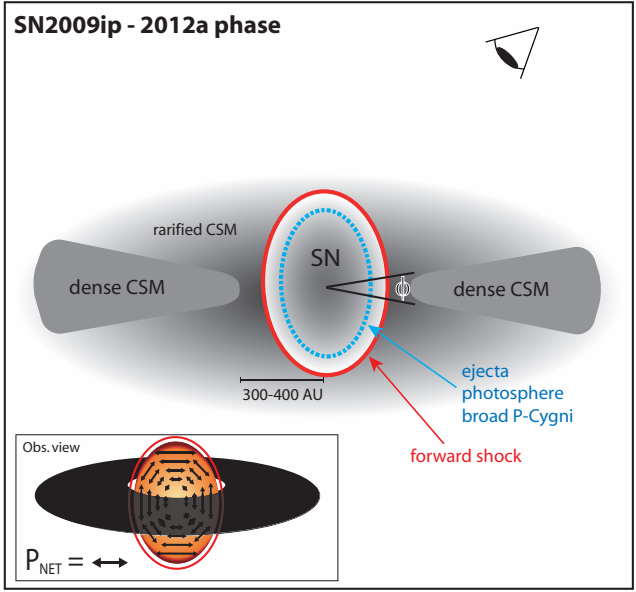}
\caption{Illustration of a potential SN/CSM configuration during the 2012a event (not to scale), before the onset of intense CSM interaction. 
The presence of broad P-Cygni lines implies a rapidly expanding photosphere. The detection of significant continuum polarization at this phase suggest an aspherical geometry for the SN photosphere and/or partial absorption by toroidal CSM (inset). The approximate orthogonality of the polarization position angle with respect to the 2012b CSM-interaction phase suggests that the SN photosphere might have a bipolar geometry. The angle $\phi$ represents the angle subtended by the CSM from the point of view of the explosion.}
\label{fig:cartoon1}
\vspace{6pt}
\end{figure}

\begin{figure}
\includegraphics[width=3.3in]{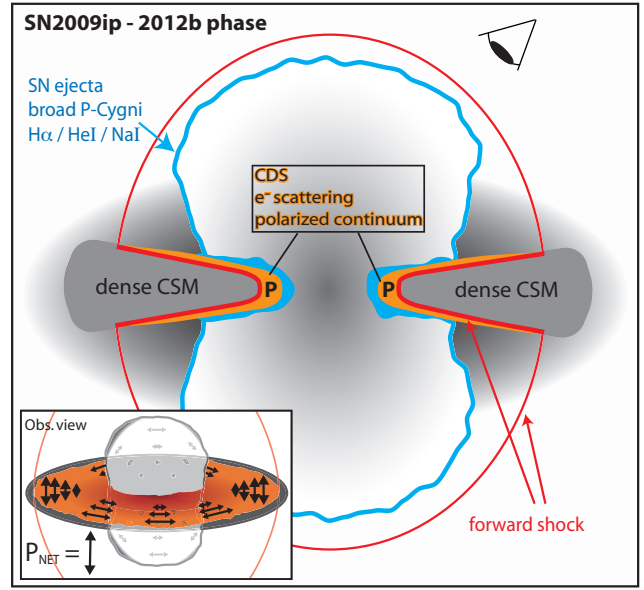}
\vspace{-40pt}
\caption{Illustration of a possible configuration for the 2012b phase, after the onset of intense CSM interaction. A toroidal distribution of dense CSM gives rise to strong shocks, electron scattering, and a luminous polarized continuum (marked by a bold-faced ``P"). At the same time, diluted yet persistent broad P-Cygni lines from 2012a indicate an underlying SN photosphere from fast ejecta that have not participated in strong CSM interaction. With respect to the observer, fast He/Na-rich ejecta in a quasi-Hubble flow could partially obscure the inner polarized continuum, resulting in the sharp polarization ``enhancements" we see at specific velocities (see Figures~\ref{fig:qu_line1} and \ref{fig:qu_line2}).}
\label{fig:cartoon2}
\end{figure}

We can estimate the true intrinsic polarization of the 2012b peak component as follows. If we rotate the $Q$--$U$ values from Figure~\ref{fig:pol_lc} such that they are aligned with the axis of symmetry at maximum polarization ($\theta=72^{\circ}$), which roughly traces the average axis of symmetry for the entirety of 2012b, then the majority of polarization during the 2012b phase is shifted onto the positive $Q$ axis, while the orthogonal 2012a polarization gets shifted onto the negative $Q$ axis. Figure~\ref{fig:rsp} shows the evolution of the resulting rotated Stokes parameters. The $Q_{\rm rot}$ curve represents variations along the main axis of symmetry, while $U_{\rm rot}$ traces deviations from the average axis of symmetry. The result implies that if the 2012a component maintains its orthogonal contribution to the total electric vector during 2012b, then the separation between the 2012a and 2012b components will approximately define the true degree of intrinsic polarization for the 2012b peak, which, therefore, could be as high as $\sim$2.6\%. In this case, comparison with the models of H{\"o}lfich (1991) would imply a more highly aspherical axial ratio of $\lesssim0.5$--0.6 for the 2012b component ($>$40\%--50\% asphericity). Considering the additional effect of a $45^{\circ}$ inclination angle, hypothetically, it is possible that the intrinsic geometry of the 2012b component is a highly flattened one. Below we discuss a plausible geometric scenario for SN~2009ip's 2012 evolution, facilitated by Figures~\ref{fig:cartoon1} and \ref{fig:cartoon2}. 
   
For the 2012b component of polarization the situation appears relatively clear. Since CSM interaction dominates the luminosity during this phase, and since the associated polarization source exhibits a rather steady and well-defined axis of symmetry, the net polarization must reflect the physical structure of the dense CSM, which in this case could represent a toroidal geometry (a disk or ring). Toroidal geometry for the interacting CSM has already been suggested for SN~2009ip by Levesque et al. (2014), based on spectral modeling and analysis of the Balmer emission lines. Those authors found that the H$\alpha$ emission during the peak of 2012b requires a total radiating area of $\sim$20,000~AU$^2$, while the Balmer decrement requires a high CSM density of $n_e>10^{13}$~cm$^{-3}$, for Case-B recombination (note that such a high density would naturally explain the absence of forbidden emission lines in SN~2009ip's spectrum). The combination of high density and large area implies a flattened geometry (i.e., a disk/ring), possibly as thin as 10 AU. Alternatively, a lower density CSM of $n_e>10^{8}$~cm$^{-3}$ in a limb-brightened spherical shell configuration was also proposed by Levesque et al. (2014) as an additional possibility, although less favourably because it requires that the optical depth of H$\beta$ be lower than for H$\alpha$, which would be unusual. Independently, SMP14 also favoured disk-like geometry for the CSM based on energetic arguments and on the persistence of broad P-Cygni lines even after the phase of intense CSM interaction had declined, which implies that a large fraction of the ejecta did not decelerate and, thus, must not have participated in strong CSM interaction (toroidal geometry would allow a large fraction of the ejected mass to bypass strong CSM interaction in directions orthogonal to the CSM plane). Our spectropolarimetric results are consistent with toroidal geometry for the CSM around SN~2009ip, and contradict models that invoke spherical distributions of CSM. In this case, the axial ratio of the polarized emission source at the peak of 2012b (constrained to $<0.7$, and probably $<$0.6; see \S4.1.2) is consistent with toroidal geometry tilted at some intermediate inclination angle with respect to our line of sight. 

The spectropolarimetric evolution is \textit{inconsistent} with some physical scenarios previously proposed as possibilities for SN~2009ip's 2012 evolution. Since it is the CSM geometry that determines the polarization properties at the peak of the interaction phase, then the earlier eruption that produced such CSM should reflect the same geometry.  Thus, the orthogonal position angle shift we observe between 2012a and the peak of 2012b is not compatible with the interpretation that these respective phases represent the launching and collision of successive shells of CSM (e.g., from pair-instability eruptions), a possibility suggested by other authors (Pastorello et al. 2013; Margutti et al. 2014), since such a scenario should result in similar axes of symmetry for the two phases, not orthogonal geometries.  Rather, the orthogonality between the 2012a and 2012b events implies a physically distinct origin for each of the associated outflows. In addition, the large shift in polarization and position angle between the 2012b peak and the subsequent bump in the declining light curve on Nov. 6--7 also suggests that the bump is the result of a polarized luminosity component that is separate from that of the 2012b peak --- for example, the SN photospheric component brightening again as the CSM interaction component fades away. This is inconsistent with the interpretation that the bump in the light curve is the result of a fluctuating central source that variably illuminates the CSM, such as activity from a central surviving star, as proposed by Martin et al. (2013).

Toroidal CSM geometry would not be unique to SN~2009ip; it has been proposed before for other SNe IIn, specifically in the  cases of SN~1988Z (Chugai \& Danziger 1994, their Model B), SN~1998S (Leonard et al. 2000), and SN~1997eg (Hoffman et al. 2008).  This type of geometry is also unsurprising in the context of SN~2009ip's stellar progenitor, since rings and tori are common phenomena around LBVs, B[e] supergiants, and other evolved massive stars (Smith, Bally, \& Walawender 2007; Smith et al. 2011a). Not at all clear, however, is the connection of such geometry to the erratic behaviour of SN~2009ip over the last decade. It is possible that dense equatorial CSM is the result of binary-influenced mass loss, as has been suggested to explain the toroidal CSM of B[e] supergiants.  A connection between such objects and SN~2009ip was speculated upon by Clark et al.~(2013) in the case of LHA~115-S~18, a B[e] supergiant exhibiting erratic long-term variability that is reminiscent of LBVs. It is thus interesting to note that the radial dimensions of B[e] disks ($10^2$--$10^3$~AU; Zickgraf et al. 1986) are comparable to the size estimates for SN~2009ip's toroidal CSM (Smith et al. 2013; SMP14). 

The origin of the polarization for the 2012a phase is less clear. Since the spectrum of SN~2009ip on Sep. 21--24 resembles that of more common core-collapse SNe, the simplest explanation might be electron scattering within the photosphere of the early SN, which has yet to impact the dense CSM lying farther out. In this case, the axial ratio of $<0.85$ and the orthogonal geometric axis with respect to the peak of the CSM interaction phase would suggest a prolate/bipolar shape for the 2012a outflow. Interestingly, bipolar geometry has been suggested previously for other cases of highly polarized core-collapse SNe (e.g., see Wang et al. 2001). Alternatively, net polarization from the 2012a phase could also arise if the near side of the soon-to-be-shocked toroidal CSM is partially blocking the waist of the SN photosphere (Figure~\ref{fig:cartoon1}, inset), which could result in a polarization axis that is orthogonal to 2012b if the polarized poles of the photosphere remain relatively unobscured. Finally, scattering of the SN photons by dust in the outer CSM could provide another source of polarized photons. After all, infrared data obtained during the 2012a phase show strong evidence for circumstellar dust at a radius of $\sim$120--220~AU (Smith et al. 2013b). We note that the strength of SN~2009ip's blue continuum detected at the same time indicates that the central source is probably not completely enshrouded by the dusty component, which is consistent with the toroidal geometry inferred for the CSM. However, if dust scattering from the dense CSM were an important source of polarized flux during 2012a, we would expect the geometric axis during this phase to reflect that of 2012b rather than being orthogonal to it. Thus, electron scattering within the SN photosphere appears to be the most plausible explanation as the dominant source of polarized photons during 2012a, with net polarization either being the result of bipolar geometry for the photosphere or partial blocking of its waist by the toroidal CSM.

\subsubsection{The imprint of high-velocity ejecta}
If strong CSM interaction is confined to a toroidal geometry, then large amounts of high-velocity SN ejecta could expand relatively unimpeded. This is already indicated by the persistence of the broad P-Cygni lines (SMP14), but could also explain some of the higher-order effects we see in the spectropolarimetric data.  For example, the sharp blueshifted enhancement in polarization associated with the He~{\sc i}/Na~D feature seen on Nov. 6--7 (Figure~\ref{fig:qu_line1}) is an indication of high-velocity material. This feature has a blueshifted velocity of 13,000~km~s$^{-1}$, while the position angle across the feature approximately follows along the main axis of symmetry of this epoch in the $Q$--$U$ plane.  The fact that the peak of the enhanced polarization feature is not coincident with the flux minimum of the absorption line implies that the enhancement cannot simply be the result of absorption of unpolarized or less polarized flux (which would increase the fractional polarization at that wavelength). Rather, the net blueshift probably results from an intervening layer of fast He-rich (or Na-rich) material that lies near the outer edge of the ejecta and is \textit{partially} absorbing the inner polarized continuum. From the point of view of the observer, such partial blocking can increase the net polarization over that range of velocities by interfering with the cancellation of orthogonal polarization vectors distributed across the polarized source (e.g., also see Kasen et al. 2003).  
The high-velocity He/Na~D-rich material was probably launched during the SN in 2012a phase, while the CSM responsible for the 2012b phase was likely to have been created by the LBV progenitor eruptions during the years prior (e.g., see Graham et al. 2014, their Figure~7).

\subsection{Consequences for kinetic energy}
A toroidal or disk-like distribution of CSM around SN~2009ip has important consequences for the kinetic energy of the explosion that is inferred from the total radiated energy measured during the 2012b CSM-interaction phase. Taking the 2012a explosion time of $-47$ days, suggested by SMP14, and the fastest ejecta velocity of 13,000~km~s$^{-1}$ indicated by the blue edge of the H$\alpha$ P-Cygni line during 2012a, implies a radial distance of $\sim$350 AU from the point of explosion to the inner edge of the dense CSM; this is consistent with the limits on the minimum radius of the dust component inferred from infrared measurements during 2012a ($>$120~AU; Smith et al. 2013b), and with the 300~AU radial estimates by Levesque et al. (2014). The latter authors showed that the CSM need only be $\sim$10~AU thick to explain the H$\alpha$ luminosity in the toroidal scenario. For a 10~AU toroidal scale height, the 300--350 AU distance from the explosion would imply a very small subtended angle of $\phi\approx\pm2^{\circ}$ for the CSM from the point of view of the explosion. Such a configuration inclined at an angle of 35--45$^{\circ}$ with respect to our line of sight (from edge-on) would appear as an ellipse on the sky having an axial ratio that is consistent with the geometric constraints placed by the value of peak polarization during 2012b (see \S4.1.2). 

For such a small subtended angle of CSM, the fraction of the total explosion solid angle ($\Omega$/4$\pi$) intercepted by CSM will be only 2--3\% ($<$10\% conservatively). This minor amount of interacting area will result in an inefficient conversion of the total kinetic energy of the SN into radiation. Thus, the $\sim10^{50}$~ergs of kinetic energy inferred from the radiated energy of the explosion by Fraser et al. (2013) and Margutti et al. (2014), both of whom assume a spherical CSM configuration in their calculations, must be a significant underestimate, by 1 dex or more. A kinetic energy of $\gtrsim10^{51}$~ergs, as suggested by SMP14, seems more likely, given the demand for aspherical CSM by spectropolarimetry. 

One must also keep in mind that the shape of the explosion is an influential factor as well. A bipolar explosion, which we have shown is a plausible geometry for the 2012a outflow, will intersect even less area of the toroidal CSM. If the material is clumpy, then the effective intersecting area could be even lower than we have estimated, implying an even lower efficiency conversion of kinetic energy into radiation. 

\section{Summary and Conclusions}
We have presented multi-epoch spectropolarimetry of the 2012 outburst of SN~2009ip, covering the 2012a and 2012b phases. Since the available data imply a very low amount of interstellar absorption and ISP $<0.2$\%, we have not attempted to remove ISP in our analysis. This could potentially lead to some small systematic errors in our results, although we expect any such changes will be minor and will not significantly influence our interpretation. 

The degree of polarization at the peak of 2012b implies an approximate axial ratio of $<$0.7 (probably $<$0.6) for the polarized CSM, which is consistent with the toroidal distribution of material suggested previously (SMP14; Levesque et al. 2014). In the future, modeling of the polarization properties of shocked toroidal CSM could reveal how well the comparison to the oblate electron-scattering atmosphere models of H{\"o}filch (1991) can characterise the axial ratio of such geometry. Details aside, the spectropolarimetric results demand substantial asphericity for the source of the 2012b luminosity and are thus inconsistent with models for SN~2009ip's 2012 explosion that invoke a spherical distribution of CSM (Fraser et al. 2013; Margutti et al. 2014). For the 2012a phase, the detection of significant polarization at a  position angle that is nearly orthogonal to 2012b is consistent with a bipolar geometry for the SN outflow or partial obscuration of a potentially more spherical SN photosphere by the toroidal CSM. In any case, the orthogonality is inconsistent with the launching and collision of successive shells of CSM (Pastorello et al. 2012) generated $\approx$40 days apart (Margutti et al. 2014), since such a scenario should produce similar axes of symmetry for the 2012a and 2012b phases, not orthogonal geometries. Furthermore, the geometric parameters constrained by spectropolarimetry imply that only a small fraction of the SN ejecta ($<$10\%) participated in strong interaction with the toroidal CSM. Therefore, earlier estimates of a $\sim10^{50}$~ergs explosion that are based on the total radiated energy, which assume spherical symmetry for the CSM (Fraser et al. 2013; Margutti et al. 2014), are likely to be substantially underestimated by an order of magnitude.

The strong similarities between SN~2009ip and SN~1993J, from the time near their peak through their decline, provide another strong line of evidence for the existence of an underlying SN photosphere from high-velocity ejecta (5000--8000~km~s$^{-1}$) that persists throughout SN~2009ip's 2012 evolution ($\sim$100 days). Such a long-lasting optically thick component indicates a mass of at least a few M$_{\odot}$ for the fast outer component of the ejecta, and the associated speeds imply a kinetic energy of $\gtrsim10^{51}$~ergs for the ejecta, as suggested by SMP14. 

After peak polarization, higher-order structure in the $Q$--$U$ plane from He-{\sc i}/Na~D indicates the presence of fast metal-rich ejecta that have overrun the dense CSM, consistent with a high-velocity flow that bypassed intense CSM interaction. Meanwhile, the weakening of polarization with declining continuum flux at late times suggests that the diminishing intensity of electron scattering is responsible for the decline in polarization. The fact that the flux and polarization drop more rapidly for SN~2009ip relative to other SNe~IIn, such as SN~1997eg (Hoffman et al. 2008) or SN~2010jl (Grant Williams, 2014, private communication), also suggests that the CSM around SN~2009ip has a compact configuration, compared to these other explosions.  

The lines of evidence presented here thus favour a scenario first proposed by Mauerhan et al. (2013), in which the luminosity of the 2012a phase of SN~2009ip was the result of a $\gtrsim10^{51}$~ergs explosion (i.e., a core-collapse SN) that subsequently plowed into an aspherical (probably toroidal) distribution of CSM during 2012b, creating the jump in luminosity and the strong polarization of the source. Nonterminal scenarios are difficult to support in light of the current body of evidence, and we would have to invent a process capable of generating ejecta having $\gtrsim10^{51}$~ergs of kinetic energy without destroying the star. Still, the best evidence of progenitor death is a vanishing of the star at late times. However, CSM interaction can persist for decades or more, and generate enough luminosity to make progenitor disappearance a difficult observable to confirm. SN~1961V is a particularly controversial example (Filippenko et al. 1995; Chu et al. 2004; Smith et al. 2011b; Kochanek et al. 2011; Van Dyk et al. 2012). Moreover, it is very plausible that the progenitor of SN~2009ip's 2012 explosion left a massive companion at the explosion site, not only because massive stars are rarely solitary (Sana et al. 2012), but because toroidal or disk-like geometry for the CSM is suggestive of binary influence. As discussed by Smith~\&~Arnett (2014), repeated brief pre-SN eruptions could result from binary interaction if the primary suddenly increases its radius during late nuclear burning stages, thereby triggering binary interaction and mass ejection. Finally, we note that strong net polarization consistently observed from SNe IIn, as a class, implies that CSM interaction is a dominanting influence on their polarization properties, and that an aspherical distribution of CSM is a commonplace feature of the stellar progenitors of these explosions.

\section*{Acknowledgements}
\scriptsize 
We thank the referee for their review of this manuscript. We thank the staffs at Lick, the MMT, and Steward Observatories for their excellent assistance. We thank S. Bradley Cenko for assisting with the Lick observations. Some observations reported here were obtained at the MMT Observatory, a joint facility of the University of Arizona and the Smithsonian Institution. This research was also based, in part, on observations made with ESO Telescopes at the La Silla Paranal Observatory under programme ID 290D-5006. We thank Hien Tran for supplying polarimetric data on SN~1993J for our comparison. J.C.M. thanks Dan Kasen at UC Berkeley for insightful discussion, and we also thank Leah N. Huk at U. Denver for helpful commentary. This research was supported by NSF grants AST-1210599 (UA), AST-1211916 (UC~Berkeley), AST-1210311 (SDSU), and AST-1210372 (U. Denver). The supernova research of A.V.F.'s group at U.C. Berkeley is also supported by Gary \& Cynthia Bengier, the Richard \& Rhoda Goldman Fund, the Christopher R. Redlich Fund, and the TABASGO Foundation.


\scriptsize 

\begin{thebibliography}{99}

\bibitem[Appenzeller et al.(1998)]{1998Msngr..94....1A} Appenzeller, I., 
Fricke, K., F{\"u}rtig, W., et al.\ 1998, The Messenger, 94, 1 

\bibitem[]{} Arnett W.D.\ 1996, Supernovae and Nucleosynthesis
  (Princeton: Princeton Univ.\ Press)

\bibitem[Benetti et 
al.(1994)]{1994A&A...285L..13B} Benetti, S., Patat, F., Turatto, M., et al.\ 1994,A\&A, 285, L13 

\bibitem[Bietenholz et al.(2003)]{2003ApJ...597..374B} Bietenholz, M.~F., 
Bartel, N., \& Rupen, M.~P.\ 2003, ApJ, 597, 374 

\bibitem[Chevalier 
\& Fransson(1994)]{1994ApJ...420..268C} Chevalier, R.~A., \& Fransson, C.\ 1994, ApJ, 420, 268 

\bibitem[Chornock et al.(2010)]{2010ApJ...713.1363C} Chornock, R., 
Filippenko, A.~V., Li, W., \& Silverman, J.~M.\ 2010, ApJ, 713, 1363 

\bibitem[Chu et al.(2004)]{2004AJ....127.2850C} Chu, Y.-H., Gruendl, R.~A., 
Stockdale, C.~J., et al.\ 2004, AJ, 127, 2850 

\bibitem[Chugai(2001)]{2001MNRAS.326.1448C} Chugai, N.~N.\ 2001, MNRAS, 
326, 1448 

\bibitem[Chugai \& Danziger(1994)]{1994MNRAS.268..173C} Chugai N.~N.,
  Danziger I.~J.\ 1994, MNRAS, 268, 173

\bibitem[Clark et al.(2013)]{2013A&A...560A..10C} Clark, J.~S., Bartlett, E.~S., Coe, M.~J., et al.\ 2013, A\&A, 560, A10 

\bibitem[]{} Dessart L., Hillier D.J., Gezari S., Basa S., \&
  Matheson T.\ 2009, MNRAS, 394, 21

\bibitem[Filippenko(1997)]{1997ARA&A..35..309F} Filippenko, A.~V.\ 1997, ARA\&A, 35, 309 

\bibitem[Filippenko et al.(1995)]{1995AJ....110.2261F} Filippenko, A.~V., 
Barth, A.~J., Bower, G.~C., et al.\ 1995, AJ, 110, 2261 

\bibitem[Foley et al.(2007)]{2007ApJ...657L.105F} Foley, R.~J., Smith, N., 
Ganeshalingam, M., et al.\ 2007, ApJL, 657, L105 

\bibitem[Foley et al.(2011)]{2011ApJ...732...32F} Foley R.~J., Berger E., 
Fox O., et al.\ 2011, ApJ, 732, 32 

\bibitem[Fossati et al.(2007)]{2007ASPC..364..503F} Fossati, L., Bagnulo, 
S., Mason, E., \& Landi Degl'Innocenti, E.\ 2007, The Future of Photometric, Spectrophotometric and Polarimetric Standardization, 364, 503 

\bibitem[Fraser et al.(2013)]{2013ApJ...779L...8F} Fraser, M., Magee, M., 
Kotak, R., et al.\ 2013a, ApJL, 779, L8 

\bibitem[]{} Fraser, M., et al.\ 2013b, MNRAS, 433, 1312

\bibitem[Graham et al.(2014)]{2014arXiv1402.1765G} Graham, M.~L., Sand, 
D.~J., Valenti, S., et al.\ 2014, arXiv:1402.1765 

\bibitem[]{} Hoffman, J.L., Leonard, D.C., Chornock, R., Filippenko,
  A.V., Barth, A.J., \& Matheson, T.\ 2008, 688, 1186

\bibitem[H{\"o}flich(1995)]{1995ApJ...440..821H} H{\"o}flich, P.\ 1995, ApJ, 
440, 821 

\bibitem[H{\"o}flich et al.(1996)]{1996ApJ...459..307H} H{\"o}flich, P., Wheeler, 
J.~C., Hines, D.~C., \& Trammell, S.~R.\ 1996, ApJ, 459, 307 

\bibitem[Kasen et al.(2003)]{2003ApJ...593..788K} Kasen, D., Nugent, P., 
Wang, L., et al.\ 2003, ApJ, 593, 788 

\bibitem[]{} Kochanek, C.~S., Szczygiel, D.~M., \& Stanek, K.~Z.\ 2012, ApJ, 758, 142

\bibitem[Kochanek et al.(2011)]{2011ApJ...737...76K} Kochanek, C.~S., 
Szczygiel, D.~M., \& Stanek, K.~Z.\ 2011, ApJ, 737, 76 

\bibitem[Leising et al.(1994)]{1994ApJ...431L..95L} Leising, M.~D., 
Kurfess, J.~D., Clayton, D.~D., et al.\ 1994, ApJL, 431, L95 

\bibitem[Leonard et al.(2000)]{2000ApJ...536..239L} Leonard D.~C.,
  Filippenko A.~V., Barth A.~J., Matheson T.\ 2000, ApJ, 536,
  239

\bibitem[Leonard et al.(2001)]{2001ApJ...553..861L} Leonard, D.~C., 
Filippenko, A.~V., Ardila, D.~R., \& Brotherton, M.~S.\ 2001, ApJ, 553, 861 

\bibitem[Leonard et al.(2002)]{2002AJ....124.2506L} Leonard, D.~C., 
Filippenko, A.~V., Chornock, R., \& Li, W.\ 2002, AJ, 124, 2506 

\bibitem[Leonard et al.(2006)]{2006Natur.440..505L} Leonard, D.~C., 
Filippenko, A.~V., Ganeshalingam, M., et al.\ 2006, Nature, 440, 505 

\bibitem[Levesque et al.(2014)]{2014AJ....147...23L} Levesque, E.~M., 
Stringfellow, G.~S., Ginsburg, A.~G., Bally, J., 
\& Keeney, B.~A.\ 2014, AJ, 147, 23 


\bibitem[Margutti et al.(2014)]{2014ApJ...780...21M} Margutti, R., 
Milisavljevic, D., Soderberg, A.~M., et al.\ 2014, ApJ, 780, 21 

\bibitem[Martin et al.(2013)]{2013arXiv1308.3682M} Martin, J.~C., Hambsch, 
F.-J., Margutti, R., et al.\ 2013, arXiv:1308.3682 



\bibitem[Mauerhan et al.(2013)]{2013MNRAS.430.1801M} Mauerhan, J.~C., Smith, N., Filippenko, A.~V., et al.\ 2013, MNRAS, 430, 1801 

\bibitem[]{} Maza, J., et al.\ 2009, CBET, 1928, 1

\bibitem[Miller et al.(1988)]{1988igbo.conf..157M} Miller, J.~S., Robinson, 
L.~B., \& Goodrich, R.~W.\ 1988, Instrumentation for Ground-Based Optical Astronomy, 157 

\bibitem[Miller&Stone1993]Miller J. S., Stone R. P. S., 1993, Lick Obs. Tech. Rep. 66. Lick Obs., Santa Cruz

\bibitem[Munari 
\& Zwitter(1997)]{1997A&A...318..269M} Munari, U., \& Zwitter, T.\ 1997, A\&A, 318, 269 

\bibitem[Ofek et al.(2013a)]{} Ofek, E.~O., et al.\ 2013a, Nature 494, 65

\bibitem[Ofek et al.(2013b)]{} Ofek, E.~O., et al.\ 2013b, ApJ, 768, 47

\bibitem[Pastorello et al.(2008)]{} Pastorello, A., et al.\ 2008, MNRAS, 389, 131

\bibitem[Pastorello et al.(2013)]{} Pastorello, A., et al.\ 2013, ApJ, 767, 1


\bibitem[Prieto et al.(2013)]{} Prieto, J.L., Brimacombe, J., Drake,
  A.J., \& Howerton, S.\ 2013, ApJ, 763, L27

\bibitem[Sana et al.(2012)]{2012Sci...337..444S} Sana, H., de Mink, S.~E., 
de Koter, A., et al.\ 2012, Science, 337, 444 

\bibitem[Schlegel(1990)]{1990MNRAS.244..269S} Schlegel, E.~M.\ 1990, 
MNRAS, 244, 269 

\bibitem[Schlegel et al.(1998)]{1998ApJ...500..525S} Schlegel, D.~J., 
Finkbeiner, D.~P., \& Davis, M.\ 1998, ApJ, 500, 525 

\bibitem[Schmidt et al.(1992)]{1992AJ....104.1563S} Schmidt, G.~D., Elston, 
R., \& Lupie, O.~L.\ 1992a, AJ, 104, 1563 

\bibitem[Schmidt et al.(1992)]{1992ApJ...398L..57S} Schmidt, G.~D., 
Stockman, H.~S., \& Smith, P.~S.\ 1992b, ApJL, 398, L57 

\bibitem[Serkowski et al.(1975)]{1975ApJ...196..261S} Serkowski, K., 
Mathewson, D.~S., \& Ford, V.~L.\ 1975, ApJ, 196, 261 

\bibitem[Smith (2014)]{2014arXiv1402.1237} Smith, N., ARA\&A, 52, in press (arXiv:1402.1237)

\bibitem[Smith 
\& Arnett(2014)]{2013arXiv1307.5035S} Smith, N., \& Arnett, D.\ 2014, ApJ, in press (arXiv:1307.5035) 

\bibitem[Smith(2006)]{2006ApJ.644..1151S} Smith N.\ 2006, ApJ, 644,
  1151

\bibitem[Smith 
\& Owocki(2006)]{2006ApJ...645L..45S} Smith, N., \& Owocki, S.~P.\ 2006, ApJL, 645, L45 

\bibitem[Smith et al.(2007)]{2007AJ....134..846S} Smith, N., Bally, J., 
\& Walawender, J.\ 2007, AJ, 134, 846 



\bibitem[Smith(2011)]{2011MNRAS.415.2020S} Smith N.\ 2011, MNRAS, 415,
  2020

\bibitem[Smith et al.(2011a)]{2011MNRAS.418.1959S} Smith, N., Gehrz, R.~D., 
Campbell, R., et al.\ 2011a, MNRAS, 418, 1959 

\bibitem[Smith et al.(2011b)]{2011MNRAS.415..773S} Smith, N., Li, W., 
Silverman, J.~M., Ganeshalingam, M., 
\& Filippenko, A.~V.\ 2011b, MNRAS, 415, 773 

\bibitem[Smith(2013)]{} Smith N.\ 2013a, MNRAS, 429, 2366 

\bibitem[Smith et al.(2013)]{2013MNRAS.tmp.2960S} Smith, N., Mauerhan, 
J.~C., \& Prieto, J.~L.\ 2014 (SMP14), MNRAS, 438, 1191 

\bibitem[]{} Smith N., \& McCray, R.\ 2007, ApJ, 671, L17

\bibitem[]{} Smith N., Vink J., de Koter A., 2004, ApJ, 615, 475

\bibitem[Smith et al.(2007)]{2007ApJ...666.1116S} Smith, N., Li, W., Foley, 
R.~J., et al.\ 2007, ApJ, 666, 1116 

\bibitem[Smith et al.(2008)]{2008ApJ...686..467S} Smith N., Chornock R., 
Li W., et al.\ 2008, ApJ, 686, 467 

\bibitem[Smith et al.(2010)]{2010AJ....139.1451S} Smith N., Miller
  A., Li W., et al.\ 2010, AJ, 139, 1451  

\bibitem[]{} Smith N., et al.\ 2011a, MNRAS, 415, 773 

\bibitem[]{} Smith N., Mauerhan, J.C., Kasliwal, M., \& Burgasser, A.\
  2013, MNRAS, 434, 2721

\bibitem[Trammell et al.(1993)]{1993ApJ...414L..21T} Trammell, S.~R., 
Hines, D.~C., \& Wheeler, J.~C.\ 1993, ApJL, 414, L21 

\bibitem[Tran et al.(1997)]{1997PASP..109..489T} Tran, H.~D., Filippenko, 
A.~V., Schmidt, G.~D., et al.\ 1997, PASP, 109, 489 

\bibitem[Van Dyk et al.(2000)]{2000PASP..112.1532V} Van Dyk S.~D., Peng 
C.~Y., King J.~Y., et al.\ 2000, PASP, 112, 1532 

\bibitem[Van Dyk et al.(1994)]{1994ApJ...432L.115V} Van Dyk, S.~D., Weiler, 
K.~W., Sramek, R.~A., Rupen, M.~P., \& Panagia, N.\ 1994, ApJL, 432, L115 

\bibitem[Van Dyk 
\& Matheson(2012)]{2012ApJ...746..179V} Van Dyk, S.~D., \& Matheson, T.\ 2012, ApJ, 746, 179 

\bibitem[Wang et al.(2001)]{2001ApJ...550.1030W} Wang, L., Howell, D.~A., 
H{\"o}flich, P., \& Wheeler, J.~C.\ 2001, ApJ, 550, 1030 

\bibitem[Wang \& Wheeler(2008)]{2008ARA&A..46..433W} Wang, L., \& Wheeler, J.~C.\ 2008, ARAA, 46, 433 

\bibitem[Zickgraf et 
al.(1986)]{1986A&A...163..119Z} Zickgraf, F.-J., Wolf, B., Leitherer, C., Appenzeller, I., \& Stahl, O.\ 1986, A\&A, 163, 119 

\end{thebibliography}
\end{document}